\documentclass{article}

\usepackage{arxiv}

\usepackage[utf8]{inputenc} 
\usepackage[T1]{fontenc}    
\usepackage{lmodern}        
\usepackage{hyperref}       
\usepackage{url}            
\usepackage{booktabs}       
\usepackage{amsfonts}       
\usepackage{nicefrac}       
\usepackage{microtype}      
\usepackage{graphicx}

\title{Big Ideas in Sports Analytics and Statistical Tools for their
Investigation}

\author{
    Benjamin S. Baumer
   \\
    Statistical \& Data Sciences \\
    Smith College \\
  Northampton, MA 01063 \\
  \texttt{\href{mailto:bbaumer@smith.edu}{\nolinkurl{bbaumer@smith.edu}}} \\
   \And
    Gregory J. Matthews
   \\
    Mathematics and Statistics \\
    Loyola University Chicago \\
  Chicago, IL 60660 \\
  \texttt{\href{mailto:gmatthews1@luc.edu}{\nolinkurl{gmatthews1@luc.edu}}} \\
   \And
    Quang Nguyen
   \\
    Statistics \& Data Science \\
    Carnegie Mellon University \\
  Pittsburgh, PA 15213 \\
  \texttt{\href{mailto:nmquang@cmu.edu}{\nolinkurl{nmquang@cmu.edu}}} \\
  }

\newlength{\cslhangindent}
\newlength{\csllabelwidth}
\newlength{\cslentryspacingunit} 
  {}%
  {\par}
\newenvironment{CSLReferences}[2] 
 {
  \setlength{\parindent}{0pt}
  \ifodd #1
  \let\oldpar\par
  \def\par{\hangindent=\cslhangindent\oldpar}
  \fi
  \setlength{\parskip}{#2\cslentryspacingunit}
 }%
 {}
\usepackage{calc}

\usepackage{amsmath}
\usepackage{booktabs}
\usepackage{longtable}
\usepackage{array}
\usepackage{multirow}
\usepackage{wrapfig}
\usepackage{float}
\usepackage{colortbl}
\usepackage{pdflscape}
\usepackage{tabu}
\usepackage{threeparttable}
\usepackage{threeparttablex}
\usepackage[normalem]{ulem}
\usepackage{makecell}
\usepackage{xcolor}
\begin{document}
\maketitle

\begin{abstract}
Sports analytics---broadly defined as the pursuit of improvement in
athletic performance through the analysis of data---has expanded its
footprint both in the professional sports industry and in academia over
the past 30 years. In this paper, we connect four big ideas that are
common across multiple sports: the expected value of a game state, win
probability, measures of team strength, and the use of sports betting
market data. For each, we explore both the shared similarities and
individual idiosyncracies of analytical approaches in each sport. While
our focus is on the concepts underlying each type of analysis, any
implementation necessarily involves statistical methodologies,
computational tools, and data sources. Where appropriate, we outline how
data, models, tools, and knowledge of the sport combine to generate
actionable insights. We also describe opportunities to share analytical
work, but omit an in-depth discussion of individual player evaluation as
beyond our scope. This paper should serve as a useful overview for
anyone becoming interested in the study of sports analytics.
\end{abstract}


\newcommand{\pkg}[1]{\textbf{#1}}

\hypertarget{introduction}{%
\section{Introduction}\label{introduction}}

Insights derived from the analysis of data have transformed the world of
sports over the last few decades. While baseball---a naturally discrete
sport with more than a century's worth of professional data---may be the
sport with the longest relationship with sports analytics, one would be
hard-pressed to identify a professional sport today in which sports
analytics is not having an impact. In basketball, analytics has driven a
shift in the conventional wisdom about shot selection. Most teams are
shooting more three-pointers, settling for fewer long two-point shots,
deploying more versatile defenders, and relying less on the strategy of
pounding the ball into the paint in an attempt to get a high-percentage
shot (\protect\hyperlink{ref-Schuhmann2021nba3point}{Schuhmann, 2021}).
In American football, teams are going for it on fourth down far more
often than in the past, a direct result of statistical analysis showing
that most teams were previously overly conservative
(\protect\hyperlink{ref-lopez2020bigger}{Lopez, 2020}). And, of course,
in baseball, teams are using defensive shifts to maximize the
probability of recording an out, encouraging hitters to improve their
launch angles, and optimizing pitcher repertoires to minimize contact
(\protect\hyperlink{ref-healey2017new}{Healey, 2017}).

These are just the most obvious examples of strategic changes that are
fueled by insights extracted from data by practitioners of sports
analytics. Similar insights are now being made in less obvious settings,
including esports (\protect\hyperlink{ref-clark2020bayesian}{Clark et
al., 2020}; \protect\hyperlink{ref-maymin2021smart}{Maymin, 2021}).
These insights come both from academia, where researchers typically use
public data to produce high-caliber, peer-reviewed scientific work, as
well as from industry, where highly-trained analysts work with with
players, coaches, and team officials to put new ideas into immediate
effect thanks to high-resolution, often proprietary data. A growing pool
of people move seamlessly between these two worlds, leading to the
formation of partnerships and the cross-pollination of ideas.

Every sport is different, with its own set of rules, strategies, methods
of data collection, number of players, and the magnitude of the role of
chance. At the same time, many sports are similar, either because one
evolved from the other, or the structure of the games share certain
attributes. Sports that are closely related historically may or may not
share common applications of analytical methods. For example, despite
belonging to the same bat-and-ball family, baseball and cricket differ
in strategies such as batting order or sacrifice plays. Conversely, with
just a few small tweaks, analytical metrics might work just as well
across sports that are unrelated and quite different. For instance, an
Elo rating could be equally valid for chess players and ice hockey
teams.

In this paper, we explore four key ideas that have widespread
applicability across many sports: the expected value of a game state
(Section \ref{sec:ev}), win probability (Section \ref{sec:wp}), measures
of team strength (Section \ref{sec:strength}), and the use of sports
betting market data (Section \ref{sec:betting}). In each case, we define
the concept mathematically, explain how it originated, and give examples
of its applications in multiple sports. Our goal is to unify the
conceptual threads, while doing some justice to the customizations
necessary to make a metric meaningful in a particular sport. We include
copious references to original works of scholarship.

Doing the work of sports analytics requires computing with data. While
the sources of sports data are too numerous to list, in Section
\ref{sec:tools} we highlight a few computational tools (including a
table of R packages) that make this kind of work possible. Section
\ref{sec:opportunities} lists several opportunities for disseminating
work publicly. We conclude in Section \ref{sec:conclusion} with a short
discussion of some ideas that are not explored in this paper. Notably,
we omit a treatment of individual player ratings for team sports, since
this concept has been covered ably in these pages by Albert
(\protect\hyperlink{ref-Albert2015}{2015}), and its inclusion would
double the length of this manuscript. We do, however, discuss individual
player ratings in the context of one-person teams (e.g., chess, tennis)
in Section \ref{sec:strength}.

We encourage readers to explore Cochran et al.
(\protect\hyperlink{ref-cochran2017oxford}{2017}) and Albert et al.
(\protect\hyperlink{ref-Albert2017handbook}{2016}) for collections of
articles in sports analytics that provide broad coverage of the field.

\hypertarget{sec:ev}{%
\section{The expected value of a game state}\label{sec:ev}}

In many sports, the first step towards an analytical understanding is
the estimation of the expected value of a game state at any given point
in it. Mathematically, we define \(X\) to be a random variable
indicating the number of points (or runs) that a team will score over
some determined amount of time (e.g., remainder of game, quarter,
period, or inning). Let \(s \in S\) be a tuple that encodes the
\emph{state} of a game. Then our task is to estimate:

\begin{equation}
  \mathbb{E}[X | s] = \sum_{x \geq 0} \Pr[X = x | s] \cdot x\,,
\end{equation}

for any state \(s \in S\), where \(\Pr[X = x | s]\) is the probability
of scoring \(x\) points given that the game is in state \(s\) and \(S\)
is the set of all possible states.

The concept of a state is easier to grasp in a sport that can be modeled
as \emph{discrete} (in the sense of
\href{https://en.wikipedia.org/wiki/Discrete-event_simulation}{discrete
event simulation}). By discrete, we mean a sport that can be easily
broken into short, distinct segments of action which are typically
summarized categorically. Each of these segments might represent a state
\(s\). For example, each pitch in baseball is either a ball or a strike.
If the ball is put in play, then there may be a complex sequence of
movements by the players, but ultimately (within a few seconds) that
sequence will end and no more action will be permitted until the next
pitch. At the beginning and end of each phase of action, we will know
definitively which team is on offense and defense, which runners are on
which bases, the score, how many outs there are, etc. Tennis could
similarly be viewed as a series of discrete actions defined by each
point. To say that a player is winning 6-2, 3-1, 40-15 and serving with
one fault committed is to characterize the state of the match. In
American football, the game can be broken down into a discrete sequence
based on each down. Contrast this to sports like lacrosse, soccer, or
any variant of hockey, which feature largely running clocks and
continuous player movement. In these sports, it is not obvious how to
break up the action into discrete chunks.

In this Section, we illustrate how the fundamental concept of the
expected value of a game state leads to compelling findings in a variety
of sports.

\hypertarget{discrete-event-analysis}{%
\subsection{Discrete event analysis}\label{discrete-event-analysis}}

First, we explore results derived from the expected value of a state in
sports where discrete event analysis is common. We draw primarily on
baseball and American football, but applications in other sports (e.g.,
tennis) are common (see for example, Kovalchik \& Reid
(\protect\hyperlink{ref-kovalchik2019calibration}{2019})).

\hypertarget{in-baseball-the-expected-run-matrix}{%
\subsubsection{In baseball, the expected run
matrix}\label{in-baseball-the-expected-run-matrix}}

In baseball, \(s\) is typically determined by two factors: the
configuration of the runners on base (there are 8 possibilities) and the
number of outs (3 possibilities). Thus, there are
\(|S| = 24 = 8 \cdot 3\) basic states of an inning in
baseball\footnote{25, if you include the absorbing state of 3 outs that
  describes the end of an inning.}, and we are often interested in the
number of runs that will be scored from some state until the end of the
inning. In this example using baseball, \(\mathbb{E}[X | s] \,\) is the
expected number of runs scored between now and the end of the inning
given that the inning is currently in state \(s\). The collection of
estimates \(\mathbb{E}[X | s]\) for all 24 states is called the
\emph{expected run matrix} \footnote{There is no inherent dimensionality
  to \(\mathbb{E}[X | s]\). The \emph{matrix} nomenclature stems from
  its values typically being displayed in \(8 \times 3\) grid. However,
  when computing with \(\mathbb{E}[X | s]\), it is most often convenient
  to treat it as a \(24 \times 1\) vector.}, and it is foundational in
baseball analytics.

\begin{figure}

{\centering \includegraphics[width=0.9\linewidth]{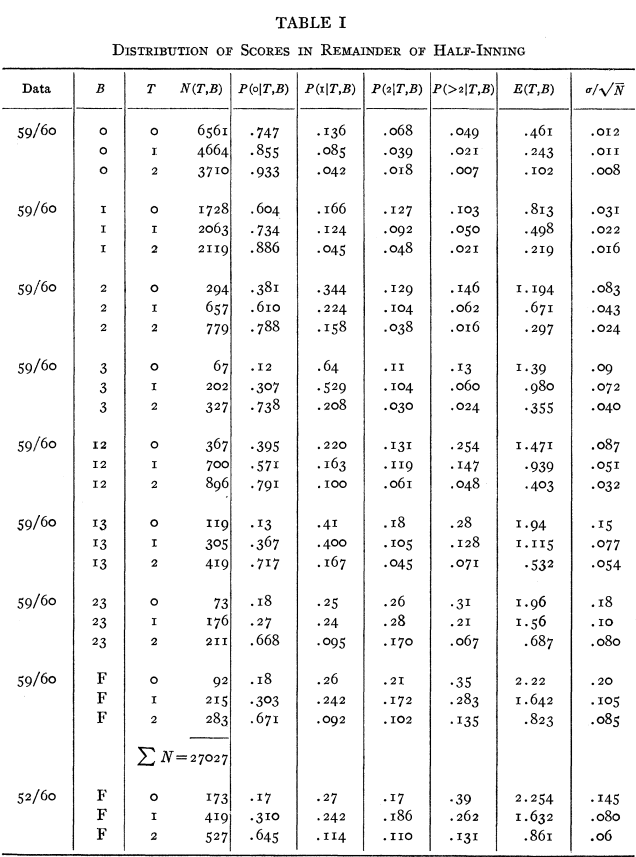} 

}

\caption{Table 1 from Lindsey's original paper. The column labeled $E(T, B)$ gives the expected run matrix as a vector, based on Lindsey's analysis of Major League Baseball data from 1959 and 1960. }\label{fig:lindsey}
\end{figure}

\begin{table}

\caption{\label{tab:lindsey-erm}George Lindsey's expected run matrix. Note how (when reading across the rows) the expected runs decrease as outs increase for the same configuration of baserunners, while (when reading down the columns) expected runs generally increase as baserunners advance. 000 means no runners on base, and 110 means runners on second and third bases. \label{tab:erm}}
\centering
\begin{tabular}[t]{lrrr}
\toprule
\multicolumn{1}{c}{} & \multicolumn{3}{c}{Out} \\
\cmidrule(l{3pt}r{3pt}){2-4}
Base & 0 & 1 & 2\\
\midrule
000 & 0.461 & 0.243 & 0.102\\
001 & 0.813 & 0.498 & 0.219\\
010 & 1.194 & 0.671 & 0.297\\
011 & 1.390 & 0.980 & 0.355\\
100 & 1.471 & 0.939 & 0.403\\
101 & 1.940 & 1.115 & 0.532\\
110 & 1.960 & 1.560 & 0.687\\
111 & 2.220 & 1.642 & 0.823\\
\bottomrule
\end{tabular}
\end{table}

Early work on this topic can be found in Lindsey
(\protect\hyperlink{ref-lindsey1963investigation}{1963}), who used
play-by-play data to compute an empirical estimate for the mean number
of runs scored in the remainder of the inning for each of these 24
possible states of an inning. This line of work led to analysis of all
types of common baseball strategies. For example, many baseball teams
elect to attempt a sacrifice bunt with a runner on first and no one out
in the inning, with the goal of moving the runner to second base, at the
cost of the batter being out. Figure \ref{fig:lindsey} shows a
reproduction of Lindsey
(\protect\hyperlink{ref-lindsey1963investigation}{1963})'s original
calculations, and Table \ref{tab:erm} shows the expected run matrix in
its most common form.

Tango et al. (\protect\hyperlink{ref-Tango2007book}{2007}) (and many
subsequent analyses) conclude that the sacrifice bunt is rarely worth
it, because most teams would be expected to score more runs with a
runner on first and no outs than they would with a runner on second and
one out.

It is worth emphasizing that the values in \(\mathbb{E}[X | s]\) are
estimates, and the precision of those estimates has many subtleties.

First, the values within the expected run matrix change over time. For
example, any estimation of the values in the expected run matrix based
on data from a high-scoring era (e.g., the early 2000s) will yield
different values than equivalent analysis in a low-scoring era. In a
high run-scoring environment, where there are many home runs, the value
of a walk may be higher, since a player who walks is more likely to
score on a subsequent home run. Conversely, in a low run-scoring
environment where hits are hard to come by, stolen bases and sacrifice
bunts may be comparatively more valuable. Thus, a careful estimate of
\(\mathbb{E}[X | s]\) would include a time parameter \(t\), indicating
when the estimate is appropriate.

Second, the characterization of \(S\) as having 24 states is only the
simplest possible. The inning, or the score of the game, or even the
weather, could be incorporated into \(S\), as those conditions might
reasonably affect the estimate of \(\mathbb{E}[X | s]\). More
definitively, the identity of the current batter, pitcher, or batter on
deck, might also affect the estimate of \(\mathbb{E}[X | s]\). Indeed,
Tango et al. (\protect\hyperlink{ref-Tango2007book}{2007}) show that
when a particularly weak-hitting batter is up (i.e., the pitcher), a
sacrifice bunt becomes a more effective strategy.

See Albert \& Bennett (\protect\hyperlink{ref-Albert2001curve}{2001})
for a fuller discussion of the use of the expected run matrix in
baseball and Marchi et al.
(\protect\hyperlink{ref-marchi2018analyzing}{2018}) for examples of how
to estimate the expected run matrix using Retrosheet data and the R
statistical computing language (\protect\hyperlink{ref-R-base}{R Core
Team, 2022}).

\hypertarget{sec:football}{%
\subsubsection{In American football, expected
points}\label{sec:football}}

The concept of estimating the value of the state of a game is easily
extended to other sports. For example, in American football, \(s\) is
determined by situational variables such as down, yardage to the next
first down, time remaining in the game, and field position.

\begin{figure}

{\centering \includegraphics[width=1\linewidth]{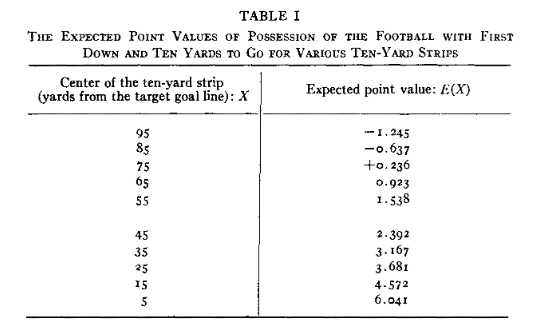} 

}

\caption{Table 1 from Carter and Machol's original paper. Note the monotonic increase in expected point values as the team gets closer to the endzone. }\label{fig:carter}
\end{figure}

The task of estimating expected points of possession in football goes
back to Carter \& Machol
(\protect\hyperlink{ref-Carter1971operations}{1971}), who estimate the
expected points for 1st and 10 plays in the NFL, given any yard line on
the football field. Due to limitations regarding the amount of data
collected, the authors divide football field into 10-yard buckets,
centered at their midpoints (e.g.~5, 15, 25, 35, etc.), before averaging
the value of the next scoring instance across the field to obtain the
expected points. Figure \ref{fig:carter} shows a reproduction of Carter
\& Machol (\protect\hyperlink{ref-Carter1971operations}{1971})'s
estimates. As expected, the estimated expected points increases
monotonically as the teams gets closer to the endzone. One limitation of
this approach is the linearity assumption, which results in a high
negative expected point value when the offensive team is 95 yards away
from the opponent's goal line.

Early work on expected point values in American football can also be
found in Carroll et al.
(\protect\hyperlink{ref-Carroll1988hidden}{1988}). In particular, the
authors consider a similar approach to Carter \& Machol
(\protect\hyperlink{ref-Carter1971operations}{1971}) and propose a
linear model for expected points in the NFL. They determine that every
extra 25 yards is associated with 2 more points scored on average for a
football team.

Other attempts at modeling expected points in football are Goldner
(\protect\hyperlink{ref-goldner2012markov}{2012}) and Goldner
(\protect\hyperlink{ref-goldner2017situational}{2017}), who propose a
Markov framework. In particular, the author considers a football drive
as an \emph{absorbing Markov chain}, consisting of distinct
\emph{absorbing states} that include touchdowns, field goals, and other
possession outcomes. An absorbing state is a link in a Markov chain from
which there are no possible transitions (i.e., it is the end of the
chain). For any given play, the expected points are calculated using the
absorption probabilities for different scoring events.

A more in-depth overview of the history of expected points in sports is
provided in Yurko et al. (\protect\hyperlink{ref-Yurko2019nflwar}{2019})
(Section 1.1). Most importantly, Yurko et al.
(\protect\hyperlink{ref-Yurko2019nflwar}{2019}) use publicly available
data provided by the \textbf{nflscrapR} package
(\protect\hyperlink{ref-R-nflscrapR}{Horowitz et al., 2020}) to model
the expected points on a play-by-play level in football. The authors
introduce a multinomial logistic regression approach, which takes into
account the current down, time remaining, yards from endzone, yards to
go, and indicators for goal down situation and whether there are less
than two minutes remaining in the half. Their model estimates the
probabilities of the following possible scoring outcomes after each
play: no score, safety, field goal, and touchdown for both the offensive
and defensive teams, all of which have a point value. The expected
points for a play can then be calculated accordingly, by summing up the
products of the scoring event point values and their associated
probabilities (see Equation 1).

In addition, Pelechrinis et al.
(\protect\hyperlink{ref-pelechrinis2019}{2019}) develop an expected
points framework in the same spirit as the previous work, but account
for the strength of the opponents in their method. They state that by
failing to account for opponent strength appropriately, about 124.8
points per team each season (or about 3.8 wins per season) are not
credited correctly. This is a substantial amount in a 16-game season.

\hypertarget{in-american-football-4th-down-strategy}{%
\subsubsection{In American football, 4th down
strategy}\label{in-american-football-4th-down-strategy}}

The concept of expected points in American football has many
applications. One of the most notable and well-studied topics is the
evaluation of 4th down strategy. There is near universal consensus in
the literature that NFL teams have been too conservative in the past
when making 4th down decisions.

Romer (\protect\hyperlink{ref-Romer2006}{2006}) examines 4th down
decisions in the NFL using expected points by focusing only on examples
from the first quarter of a game (to avoid issues with end-of-half and
end-of-game decision making). They concluded that teams don't go for it
enough if teams are trying to maximize their probability of winning the
game.

Numerous other papers (see Lopez
(\protect\hyperlink{ref-lopez2020bigger}{2020}) for details) use the
analysis of the expected number of points to improve fourth down
strategy. In addition to Romer
(\protect\hyperlink{ref-Romer2006}{2006}), later work by Yam \& Lopez
(\protect\hyperlink{ref-Yam2019}{2019}) uses win probability (see
Section \ref{sec:wp}), rather than expected points, and a causal
inference framework to reach similar conclusions that NFL teams are too
conservative in going for it on 4th down. In addition, they estimate
that a better strategy would be worth about 0.4 wins per season on
average, a substantial amount comparable to the effect size reported by
Pelechrinis et al. (\protect\hyperlink{ref-pelechrinis2019}{2019})
above.

Lopez (\protect\hyperlink{ref-lopez2020bigger}{2020}) presents an
introduction to NFL tracking data, and examines 4th down behavior as an
example of the type of problem that can be more thoroughly studied with
the increase in granularity of the tracking data over traditional NFL
data. In the past, when looking at down and distance data to study
whether NFL coaches are making good decisions about whether to ``go for
it'' or punt on 4th down, the distance data is only a rounded
approximation of the true distance ``to go'' (i.e.~1 yard, 1 foot, and 1
inch will all be recorded as 4th and 1. In fact, anything up to 2 yard
will recorded as 4th and 1 (\protect\hyperlink{ref-lopezDM}{Lopez,
2022}). However, a coach on the field during a game will be able to
clearly see the difference between 1 inch and 1 yard, and this
information will factor into their decision making. With tracking data,
the ``to go'' distance can be much more accurately assessed and
therefore evaluation of 4th down coaching decisions can now account for
this ``extra'' information that is available to a coach on the field of
play, but not recorded in traditional NFL data. Many past analyses of
the decision to go for it or not on 4th down conclude that coaches in
the NFL are too conservative in their decision making. Lopez
(\protect\hyperlink{ref-lopez2020bigger}{2020}) also concludes that
coaches are too conservative on 4th down decision making, but notes
further that past estimates of the magnitude of how conservative coaches
are on 4th down may be overstated due to the way in which to go yardage
was recorded only approximately in the past.

\hypertarget{other-applications-of-expected-points-in-american-football}{%
\subsubsection{Other applications of expected points in American
football}\label{other-applications-of-expected-points-in-american-football}}

Researchers have also applied the notion of expected points to
investigate other aspects of the game of football, including quarterback
performance and coaching decisions.

For quarterback evaluation, White \& Berry
(\protect\hyperlink{ref-White2002}{2002}) present a tiered logistic
regression method that can be, in general, applied to any regression
setting with a polychotomous response. Using this technique, they
estimate the value of NFL plays using a simple expected points model
with down, yards to go, and yards to goal as predictors. Accordingly,
the model results are utilized to obtain ratings and rankings for NFL
passers.

Alamar (\protect\hyperlink{ref-Alamar2010}{2010}) implements an expected
points framework to examine play calling in the NFL. However, rather
than assessing each play on its own, they evaluate the play in the
context of the drive. Based on play-by-play data from 2005 through 2008,
they determine that teams are under-utilizing passing plays in some
situations.

Another application of expected points is to evaluate kickoff decisions
made by football coaches, as demonstrated by Urschel \& Zhuang
(\protect\hyperlink{ref-urschel2011}{2011}). Specifically, they look at
surprise on-sides kicks versus regular kickoffs and the decision to
accept a touchback versus returning the kickoff. Using data from the
2009 NFL season, they conclude, as many have, that coaches in the NFL
tend to make conservative decisions.

\hypertarget{continuous-event-analysis}{%
\subsection{Continuous event analysis}\label{continuous-event-analysis}}

Even in sports where the concept of a state is more difficult to define,
the value of a possession can be estimated with the help of tracking
data. Over the past decade or so, professional sports leagues have
collected tracking data which record the locations of all players and
the ball (or puck) throughout a game. This high-resolution data allows
researchers to produce advanced analyses of the captured spatiotemporal
information and better understand the game. This is a great leap forward
from older resources such as traditional box-score results and
play-by-play data.

\hypertarget{in-basketball-expected-point-value}{%
\subsubsection{In basketball, expected point
value}\label{in-basketball-expected-point-value}}

In basketball, Cervone et al.
(\protect\hyperlink{ref-cervone2014pointwise}{2014}) and Cervone et al.
(\protect\hyperlink{ref-cervone2016multiresolution}{2016}) introduce
expected possession value (EPV) as a means toward an assessment of a
player's on-court performance. This metric is a continuous-time estimate
of the expected number of points for the offensive team on a given
possession using player and ball locations. The EPV takes into account
all possible outcomes (a shot attempt, a pass, etc.) for a given player
with the ball, with different weights being assigned to each decision.
The computation of the EPV statistic is done using a (technically
discrete) Markov model conditioned on spatial locations. Consequently,
the authors derive a metric called EPV-Added (EPVA), measuring a
player's EPV contribution in a given situation relative to a
league-average player.

\begin{figure}

{\centering \includegraphics[width=1\linewidth]{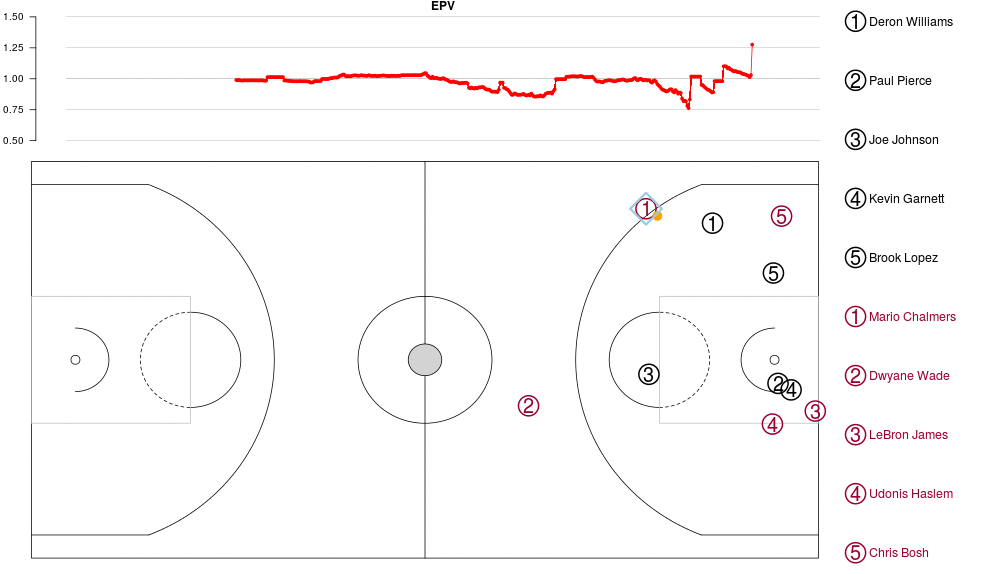} 

}

\caption{Player locations and estimated EPV for a possession during the Miami Heat (red) vs. Brooklyn Nets (black) NBA game on November 1, 2013. The captured moment is when Miami's Mario Chalmers just releases a three-point shot, which ends up as a missed field goal. Figure created by Cervone, et al. \url{https://github.com/dcervone/EPVDemo}}\label{fig:epv-demo}
\end{figure}

A demonstration of the EPV model presented in Cervone et al.
(\protect\hyperlink{ref-cervone2016multiresolution}{2016}) is available
at \url{https://github.com/dcervone/EPVDemo}. Figure \ref{fig:epv-demo}
illustrates how the provided tracking data informs the evolution of EPV
throughout the play. It displays a snapshot of a possession during the
NBA regular season matchup between the Miami Heat and the Brooklyn Nets
on November 1, 2013. Miami is the team on offense in this possession,
whose outcome is a 26-foot three-point miss by Mario Chalmers. The plot
consists of two elements: 1) (bottom) the player locations on the court
at a particular moment in this possession: when the ball just left
Chalmers's hands, and 2) (top) a line graph showing how the EPV changes
continuously throughout the play until the three-point attempt. For this
possession, the estimated EPV for the Miami Heat reaches its peak at
\(1.276\) points at the moment the shot is taken.

Note that Miami starts the play with an EPV of approximately 1.0 points,
which indicates their implied average points per possession. Chalmers'
shot is worth 3 points, so the EPV of 1.276 points implies that the
model estimate of the probability of Chalmers making this shot is
42.5\%. A breakthrough in this work is that this estimate is conditional
on the locations of the other 9 players on the basketball court.

Another framework for estimating expected points in basketball is
proposed by Sicilia et al.
(\protect\hyperlink{ref-Sicilia2019DeepHoops}{2019}). The authors offer
a different point of view on expected points, where they first consider
a classification model which returns the probabilities for whether a
player would commit a foul (shooting and non-shooting), turnover, or
attempt a shot. The values associated with each of those ``terminal
actions'' are then used to compute the expected points within a
basketball play.

See also Bornn et al. (\protect\hyperlink{ref-bornn2017studying}{2017})
for more information on how tracking data have enabled advanced
statistical analyses of basketball in recent years. The strategy of
maximizing expected points in basketball has led directly to the
proliferation of three-point shooting in the NBA.

\hypertarget{in-american-football-yards-gained}{%
\subsubsection{In American football, yards
gained}\label{in-american-football-yards-gained}}

In Section \ref{sec:football}, we discussed advances in American
football analytics based on discrete game states defined by down, yards
to first down, field position, etc. The advent of player tracking data
makes it possible to analyze American football using continuous states.
For example, Yurko et al. (\protect\hyperlink{ref-Yurko2020going}{2020})
use tracking data provided by the
\href{https://operations.nfl.com/gameday/analytics/big-data-bowl/past-big-data-bowl-recaps/}{2019
NFL Big Data Bowl} (see Section \ref{sec:opportunities}) to model the
expected yards gained for a ball-carrier during the course of a play. As
an extension to pre-existing approaches, the authors use conditional
density estimation to obtain a probability distribution for the number
of yards gained during the play, rather than only producing a single
estimate for the expected yards gained. Accordingly, the probability of
various types of outcomes at the end of a play such as a touchdown or a
first-down gain can be computed from the distribution of the end-of-play
yard line.

Expected point value is also the main component of a novel NFL
quarterback evaluation metric introduced by Reyers \& Swartz
(\protect\hyperlink{ref-Reyers2021quarterback}{2021}). The authors take
advantage of player tracking data to account for different passing and
running options on the football field that are available to the
quarterback. The expected points and probabilities associated with the
possible quarterback options are estimated using the method of ensemble
learning via stacking.

\hypertarget{in-other-sports}{%
\subsubsection{In other sports}\label{in-other-sports}}

The notion of expected possession value has also been extended to
association football (soccer). Fernandez et al.
(\protect\hyperlink{ref-fernandez2021framework}{2021}) implement deep
learning methods to examine the instantaneous expected value of soccer
possessions. This approach considers passes, ball drives, and shots in
soccer as the main set of actions used to compute the EPV metric. Many
applications can be derived from this framework, including predicting
which footballer on the pitch is most likely to receive the next pass
from the current on-ball player.

Macdonald (\protect\hyperlink{ref-macdonald2012expected}{2012}) uses
expected goals to evaluate ice hockey players, but does not have access
to player tracking data necessary to evaluate possessions. Kumagai et
al. (\protect\hyperlink{ref-kumagai2021hockey}{2021}) offer an EPV
metric for ice hockey via a Bayesian space-time framework.

\hypertarget{optimal-strategies-that-dont-maximize-expected-points}{%
\subsection{Optimal strategies that don't maximize expected
points}\label{optimal-strategies-that-dont-maximize-expected-points}}

Earlier in Section \ref{sec:ev}, we defined the expected value of a
possession based on the state \(s\) of the game in terms of the expected
number of points (runs) \(X\) that would be scored in the remainder of
some period of time. We then showed how this value could be used to
analyze the relative effectiveness of certain strategies, with the
simple idea that strategies that yield higher expected values are
preferable. Generally, the goal of any sport is to score more points
than the other team, which most often means trying to score as many
points as possible, leading to a general strategy of maximizing expected
points. However, there are situations in which maximizing the number of
expected points is \emph{not} the desired strategy.

For example, in the bottom of the ninth inning of a tied baseball game,
the optimal strategy for winning the game is maximizing the probability
of scoring \emph{at least one run}, which may differ from the strategy
of maximizing expected runs. If we let \(U\) be the set of all
strategies, then we assert that it is not always the case that the
strategy \(u \in U\) that maximizes the expected number of points will
maximize the probability of winning: \[
  \underset{u \in U}{\arg \max \,}{\Pr[X > 0 | s, u]} \neq \underset{u \in U}{\arg \max \,}{\mathbb{E}[X | s, u]} \,.
\] Consider the situation where runners are on first and third base, and
the score is tied in the bottom of the ninth inning with no one out.
Information derived from Table \ref{tab:erm} reveals that the expected
number of runs scored in the remainder of the inning is 1.94 runs, while
the probability of scoring zero runs is 0.13. The defense is in a tight
spot, facing an 87\% probability of losing the game. However, by walking
the hitter to load the bases, they create the opportunity to force the
lead runner at home and thus reduce the chances of scoring to 82\%, even
though they raise the expected number of runs scored to 2.22. In this
case, the \emph{defensive} team is wise to pursue the strategy of
maximizing the expected number of runs scored, because it
\emph{minimizes} the probability of scoring at least one run.

Maximizing the probability of scoring is optimal in any sudden-death
situation, which has (but currently does not) included overtime in
American football (\protect\hyperlink{ref-martin2018markov}{Martin et
al., 2018}).

The situation gets even more interesting when teams modify both their
offensive and defensive strategies simultaneously. For example, in
hockey teams will often pull their goalie when trailing in the final
period. This strategy severely weakens their defense but strengthens
their offense. The hope is to score a quick goal to get back in the
game, but the risk is falling further behind. Beaudoin \& Swartz
(\protect\hyperlink{ref-beaudoin2010strategies}{2010}) show that NHL
coaches do not always employ the optimal strategies, usually by waiting
too long to pull their goalies. Skinner
(\protect\hyperlink{ref-skinner2011scoring}{2011}) develops a general
framework for these desperation strategies, which include the onside
kick in American football, pulling the infield and/or outfield in
baseball, and of course, the fabled Hack-a-Shaq strategy in basketball.

\hypertarget{sec:wp}{%
\section{Win probability}\label{sec:wp}}

A related, but different concept to expected points is the notion of
\emph{win probability}. Win probability is simply an estimate of the
probability that a team will win the game, given its current state
\(s\). Extending the mathematical framework we defined in Section
\ref{sec:ev}, let \(W_i\) be a binary random variable that indicates a
win for team \(i\). Then, \[
  \Pr[W_i | s] \,,
\] is the win probability for team \(i\) in the state \(s\).

This win probability is closely related to the expected value of a
state. Albert (\protect\hyperlink{ref-Albert2015}{2015}) defines the win
probability as: \[
  \Pr[W_i | s] = \sum_{X \geq 0} \Pr[X | s] \cdot \Pr[W_i | X, s] \,,
\] where \(\Pr[W_i | X, s]\) is the probability that team \(i\) will win
the game given that they score \(X\) points from state \(s\).

Win probability is easily extended to provide a measure of the impact of
sports plays and individual player contributions, as discussed in Albert
(\protect\hyperlink{ref-Albert2015}{2015}). Given its popularity, recent
books on sports analytics often dedicate multiple chapters entirely to
win probability. These include Albert \& Bennett
(\protect\hyperlink{ref-Albert2001curve}{2001}), Schwarz
(\protect\hyperlink{ref-Schwarz2005numbers}{2004}), Tango et al.
(\protect\hyperlink{ref-Tango2007book}{2007}), Albert et al.
(\protect\hyperlink{ref-Albert2017handbook}{2016}), and Winston et al.
(\protect\hyperlink{ref-Winston2022}{2022}).

In this section, we discuss notable previous work on win probability in
baseball, American football, basketball, and several other sports.

\hypertarget{baseball}{%
\subsection{Baseball}\label{baseball}}

The notion of win probability in baseball goes back to at least as early
as Lindsey (\protect\hyperlink{ref-Lindsey1961}{1961}), who calculates
the expected win probability after each inning based on the distribution
of runs scored in each inning. Inspired by Lindsey
(\protect\hyperlink{ref-Lindsey1961}{1961})'s work, Mills \& Mills
(\protect\hyperlink{ref-Mills1970}{1970}) utilize win probability to
introduce Player Win Average (PWA), a measure of a player's contribution
to the game outcome. In particular, PWA is computed as \[
PWA = \frac{Win \ Points}{Win \ Points + Loss \ Points},
\] where the win and loss points represent how much the player
positively and negatively impacts their team's probability of winning
after each play. In effect, the win points are the sum of the changes in
\(\Pr[W_i | s]\) from one state to the next.

Additionally, a mathematical model for estimating win probability in
baseball is presented in Tango et al.
(\protect\hyperlink{ref-Tango2007book}{2007}). The authors use Markov
chains to look at win expectancy throughout the course of a baseball
game. This approach considers different states of the game such as base,
inning, outs and score, and outputs win probabilities accordingly.

See Albert (\protect\hyperlink{ref-Albert2015}{2015}) for a more
detailed historical overview of the use of win probability in baseball.

\hypertarget{american-football}{%
\subsection{American football}\label{american-football}}

In recent years, a number of statistical methods have been used to build
well-calibrated win probability models in American football. These are
flexible algorithms that have high predictability, can account for
nonlinear interactions between the explanatory variables, require few
assumptions, and produce feature importance scores.

Lock \& Nettleton (\protect\hyperlink{ref-Lock2014}{2014}) implement a
random forest framework to provide a win probability estimate before
each play in a football game. Covariates included in this tree-based
method are the current down, score differential, time remaining,
adjusted score, point spread, number of timeouts remaining for each
team, total points scored, current yard line, and yards to go for a
first down. According to this model, the difference in score between the
two teams is the most important feature for predicting win probabilities
at any given moment in an NFL game.

In addition, Yurko et al.
(\protect\hyperlink{ref-Yurko2019nflwar}{2019}) estimate win probability
in the NFL using a generalized additive model (GAM), as part of the
\textbf{nflscrapR} package (\protect\hyperlink{ref-R-nflscrapR}{Horowitz
et al., 2020}) and nflWAR framework. This model takes into account the
estimated expected points obtained from the model described in Section
\ref{sec:ev}, along with other predictors for time, current half, and
timeouts. The two win probability frameworks proposed by Lock \&
Nettleton (\protect\hyperlink{ref-Lock2014}{2014}) and Yurko et al.
(\protect\hyperlink{ref-Yurko2019nflwar}{2019}) were also implemented in
Yam \& Lopez (\protect\hyperlink{ref-Yam2019}{2019}) with minimal
modifications. Specifically, the authors combined both approaches to
estimate the win probability for each play, with an overall goal of
assessing fourth down decision-making in American football.

A vital highlight of Yurko et al.
(\protect\hyperlink{ref-Yurko2019nflwar}{2019})'s win probability model
is that it is fully reproducible and uses publicly available data. One
of Yurko et al. (\protect\hyperlink{ref-Yurko2019nflwar}{2019})'s goals
was also to encourage researchers to ``use, explore, and improve upon
our work,'' which ultimately inspired \pkg{nflfastR} (Carl \& Baldwin
(\protect\hyperlink{ref-R-nflfastR}{2022})), now considered the
successor to \textbf{nflscrapR}.

\begin{figure}

{\centering \includegraphics{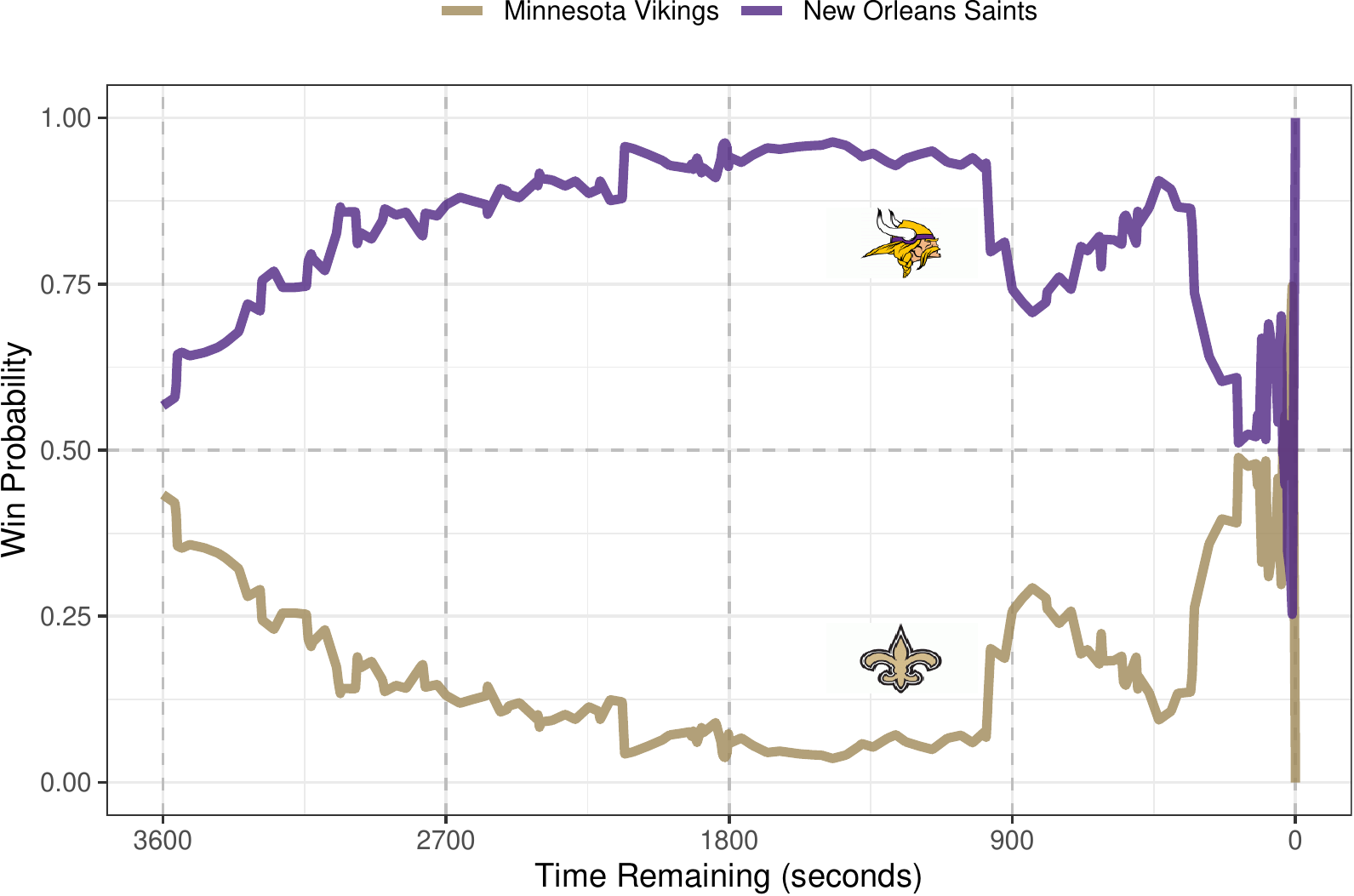} 

}

\caption{Win probability graph for New Orleans Saints vs. Minnesota Vikings in the 2017--18 NFL Playoffs.}\label{fig:nflfastr-wp}
\end{figure}

Figure \ref{fig:nflfastr-wp} shows a win probability graph for the 2018
NFL Playoffs Divisional Round matchup between the New Orleans Saints and
the Minnesota Vikings on January 14, 2018. We obtain the estimated
probability of winning for each team using the \pkg{nflfastR} R package,
which implements a gradient boosting model via the \pkg{xgboost} library
(Chen et al. (\protect\hyperlink{ref-R-xgboost}{2022})) for estimating
win probabilities. Minnesota was leading throughout the first three
quarters of the game, having win probabilities of 0.869, 0.941, and
0.742 at the end of the first, second, and third quarters, respectively.
The win probabilities get close to parity late in the fourth quarter,
when the Saints took the lead with 3:01 left in the game. The last play
of this game---famously known as the
\href{https://en.wikipedia.org/wiki/Minneapolis_Miracle}{Minneapolis
Miracle}---resulted in a drastic swing in win probabilities for both
teams. With 10 seconds remaining in the game, the Vikings begin the
final possession with a 25.3\% chance of winning. Their probability
increased to a perfect 1 when Stefon Diggs scored a game-winning 61-yard
receiving touchdown as the game clock expired.

\hypertarget{basketball}{%
\subsection{Basketball}\label{basketball}}

Stern (\protect\hyperlink{ref-Stern1994}{1994}) provides an
investigation of in-game win probability and the scoring process in
basketball using a Brownian motion model. Let \(p(l, t)\) represent the
win probability for the home team given an \(l\)-point lead after \(t\)
seconds of game time. The model introduced by Stern
(\protect\hyperlink{ref-Stern1994}{1994}) is a probit regression model,
which provides an estimate for \(p(l, t)\). Specifically, \[
p(l, t) = \Phi\left(\frac{l + (1-t)\mu}{\sqrt{(1-t)\sigma^2}}\right)
\,.
\] Here, a Brownian motion process with drift \(\mu\) points advantage
for the home team and variance \(\sigma^2\) is used to model the score
difference between the home and away teams.

On a related note, Deshpande \& Jensen
(\protect\hyperlink{ref-Deshpande2016}{2016}) extend Stern
(\protect\hyperlink{ref-Stern1994}{1994})'s framework by applying it in
a Bayesian setting. Deshpande \& Jensen
(\protect\hyperlink{ref-Deshpande2016}{2016}) propose a Bayesian linear
regression model to assess the impact of individual players on their
team's chance of winning at any given time of a basketball game. This
model assumes independence of observations and constant variability in
win probability.

Moreover, McFarlane (\protect\hyperlink{ref-McFarlane2019}{2019}) uses
logistic regression to estimate win probability for evaluating
end-of-game decisions in the NBA. The approach takes into account the
remaining game time, score difference, and point spread. This win
probability model is then applied to the calculation of the End-of-game
Tactics Metric (ETM), measuring how the chance of winning a basketball
game differs between the optimal and on-court actual decisions.

\hypertarget{other-sports}{%
\subsection{Other sports}\label{other-sports}}

The idea of win probability is also applied to other sports, with a
diverse range of statistical techniques being used to estimate the
probability of winning for a player or team. Brenzel et al.
(\protect\hyperlink{ref-Brenzel2019}{2019}) use three-dimensional Markov
models to estimate win probability throughout a curling match. In
particular, the authors propose both homogeneous and heterogeneous
Markov models for estimating the chance of winning in curling, with
different independence assumptions on the relationship between
performance and the current state of the game. In esports, Maymin
(\protect\hyperlink{ref-maymin2021smart}{2021}) relies on logistic
regression to build a well-calibrated in-game win probability model for
each specific moment during a game of League of Legends. Moreover, Guan
et al. (\protect\hyperlink{ref-Guan2022}{2022}) develop an in-game win
probability model for the National Rugby League using functional data
analysis. In this approach, the rugby play-by-play event data are
treated as functional, and the win probability is expressed as a
function of the match time.

\hypertarget{sec:strength}{%
\section{Team strength}\label{sec:strength}}

A third crucial idea in sports analytics is the estimation of team
strength. First, we briefly introduce a simple method for estimating
team strength based on win-loss record. Next, we detail three other more
sophisticated methods for estimating team strength in sports through
pairwise evaluations. The methods in this Section apply equally well to
multiplayer teams and single-player teams.

The impetus for all methods for estimating team strength is the
realization that win-loss records are a noisy measure of team strength.
As binary outcomes, and with all sports (except perhaps chess) involving
some element of chance, wins and losses carry some signal of team
strength, but we can do better.

\hypertarget{sec:wpct}{%
\subsection{Expected winning percentage}\label{sec:wpct}}

A simple method for estimating team strength that has become popular in
sports analytics is expected winning percentage---often called
Pythagorean expectation---developed by James
(\protect\hyperlink{ref-james2003new}{2003}). Later, Miller
(\protect\hyperlink{ref-miller2007derivation}{2007}) derived the formula
as a consequence of assuming that runs (in baseball) are generated by
two independent Weibull processes.

Expected winning percentage is just: \[
  \widehat{wpct} = \frac{X^\alpha}{X^\alpha + Y^\alpha} \,,
\] where \(X\) is the number of points (runs) that a team has scored,
and \(Y\) is the number of points (runs) that they have allowed, over
some specified time period. James's work was originally in baseball, and
he posited the value of \(\alpha = 2\). The resemblance to the formula
for computing the length of the hypotenuse in a right triangle provides
the nod to Pythagoras.

\begin{figure}

{\centering \includegraphics{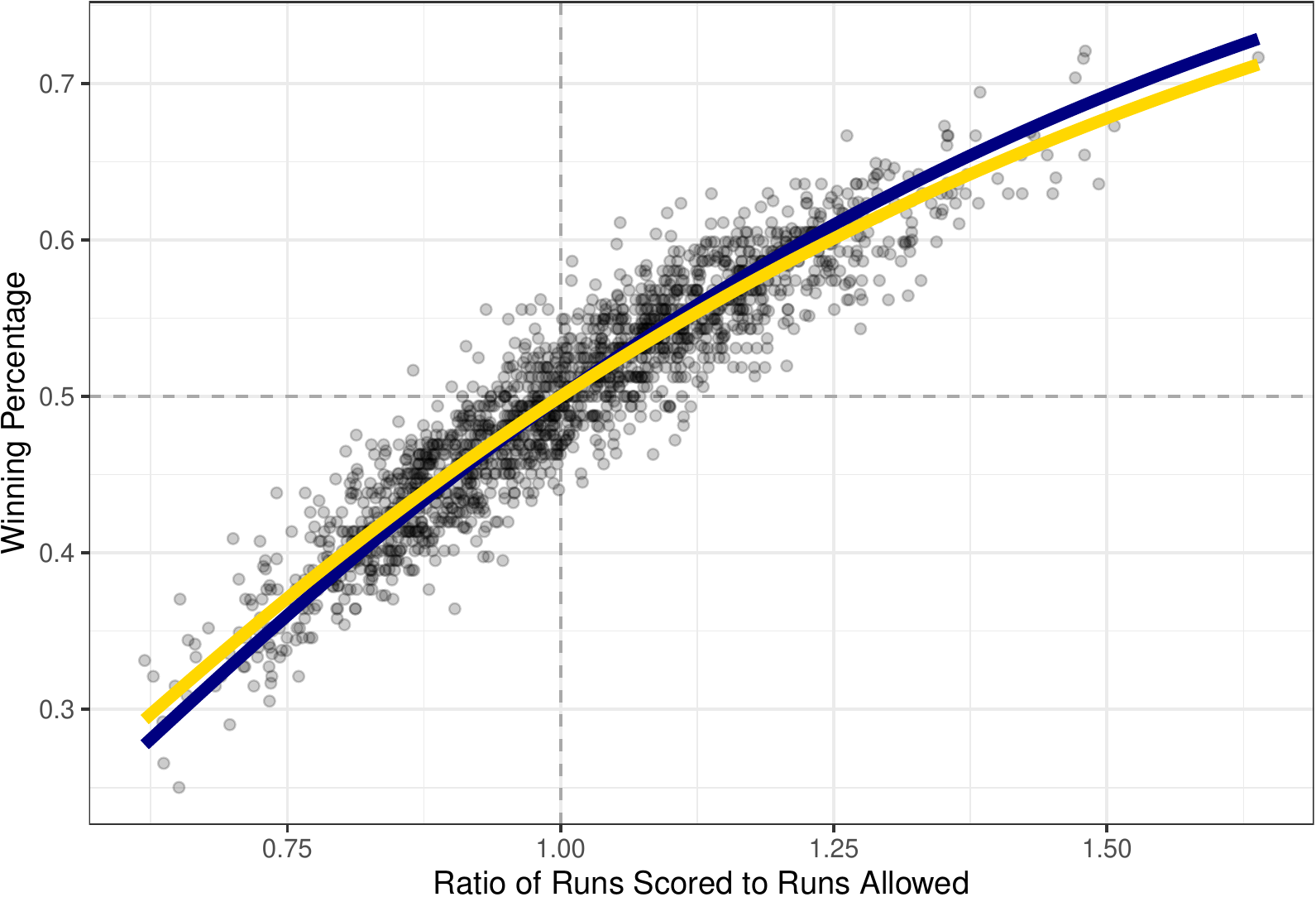} 

}

\caption{Winning percentages vs. runs scored and runs allowed in baseball, 1954--2021. The navy line represents the expected winning percentage model posited by James, with the exponent value of 2. The gold line shows the same model with an optimal exponent of 1.84.}\label{fig:pythag}
\end{figure}

Subsequent analysts have tried to find the optimal value of \(\alpha\)
for various time periods. This can be done with a few lines of code,
after observing that \[
  \frac{X^\alpha}{X^\alpha + Y^\alpha} = \frac{1}{1 + (Y/X)^\alpha}
\] and fitting a non-linear model (see similar discussion in Baumer et
al. (\protect\hyperlink{ref-baumer2021mdsr}{2021})). Figure
\ref{fig:pythag} illustrates the quality of the fit in Major League
Baseball since 1954, where the optimal value of \(\alpha\) is 1.84.

Many authors have fit expected winning percentage models to other
sports---too many to cite here. See, for example, Hamilton
(\protect\hyperlink{ref-hamilton2011extension}{2011}) for association
football (soccer), Caro et al.
(\protect\hyperlink{ref-caro2013testing}{2013}) for Division I college
football, and notably, future NBA general manager Daryl Morey for
basketball (\protect\hyperlink{ref-dewan1993stats}{Dewan \& Zminda,
1993}).

\hypertarget{bradley-terry-models}{%
\subsection{Bradley-Terry models}\label{bradley-terry-models}}

Perhaps the most widely-used probability model for predicting the
outcome of a paired comparison is the Bradley-Terry model (BTM)
(\protect\hyperlink{ref-Bradley1952}{Bradley \& Terry, 1952}). For a
pair of players (or teams) \(i\) and \(j\), let \(\Pi_{ij}\) denote the
probability that \(i\) is preferred to \(j\). Then the BTM is a logistic
regression model with parameters \(\beta_i, \beta_j\) such that \[
\log\left(\frac{\Pi_{ij}}{\Pi_{ji}}\right) = \beta_i - \beta_j \,.
\] Here, \(\exp{(\beta_i)}\) is often viewed as a representation of team
\(i\)'s ability.

The BTM can be implemented in R via the \pkg{BradleyTerry2} package
(\protect\hyperlink{ref-R-BradleyTerry2}{Turner \& Firth, 2020}). As an
example, we consider the data given in Agresti
(\protect\hyperlink{ref-agresti2018introduction}{2018}) (page 247) on
tennis results from 2014--2018 for five men's professional players:
Novak Djokovic, Roger Federer, Andy Murray, Rafael Nadal, and Stan
Wawrinka. We fit a BTM to estimate the win probability for each pair of
players and obtain a ranking for this group of five.

\begin{table}[!h]

\caption{\label{tab:unnamed-chunk-5}The estimated abilities (with standard errors) for each tennis player, relative to Wawrinka, obtained from the fitted Bradley-Terry model. \label{tab:tennisbtm}}
\centering
\begin{tabular}[t]{lrr}
\toprule
Player & Ability & SE\\
\midrule
Djokovic & 1.176 & 0.500\\
Federer & 1.136 & 0.511\\
Wawrinka & 0.000 & 0.000\\
Nadal & -0.062 & 0.515\\
Murray & -0.569 & 0.568\\
\bottomrule
\end{tabular}
\end{table}

Table \ref{tab:tennisbtm} shows the estimated coefficients of the fitted
BTM. According to the abilities, between 2014 and 2018 the players are
ranked as follows: 1) Djokovic, 2) Federer, 3) Wawrinka, 4) Nadal, 5)
Murray. In addition to an ordering, the magnitude of the coefficients in
Table \ref{tab:tennisbtm} provide a measure of relative strength.

To obtain win probabilities, as an illustration, for the Federer-Nadal
matchup, an estimate for the probability of a Federer victory is: \[
\hat\Pi_{24} = \frac{\exp(\hat\beta_2 - \hat\beta_4)}{1 + \exp(\hat\beta_2 - \hat\beta_4)} = \frac{\exp(1.136 + 0.062)}{1 + \exp(1.136 + 0.062)} = 0.768 \,.
\]

\hypertarget{sec:elo}{%
\subsection{Elo ratings}\label{sec:elo}}

Another widely known tool for measuring team strength is the Elo rating
system (\protect\hyperlink{ref-Elo1978}{Elo, 1978}), which was
originally developed for chess. Given two players \(i\) and \(j\) with
unknown ratings \(R_i\) and \(R_j\), the probability \(\Pi_{ij}\) of
\(i\) beating \(j\) is defined as \[
\Pi_{ij} = \frac{1}{1 + K^{(R_j - R_i)/400}} \,.
\] In this formula, \(K\) is commonly known as the \(K\)-factor, or
development coefficient. The International Chess Federation (FIDE) uses
\(K=10\) for players with any previously achieved rating of at least
2400. Finally, \(K=40\) is given to new players with under 30 games
played, and players under the age of 18 with rating less than than 2300
(\protect\hyperlink{ref-fide2022rating}{FIDE, 2022}).

Another interpretation for \(\Pi_{ij}\) is the expected score of the
game for player \(i\). The scores of 0, 0.5, and 1 are associated with
three possible game outcomes loss, tie, and win, respectively. After a
game, the updated Elo rating \(R^*_i\) for player \(i\) is \[
R^*_i = R_i + K(S_i - \Pi_{ij}) \,,
\] where \(S_i \in \{0, 0.5, 1\}\). When a tournament concludes, a
post-tournament rating is obtained for each player based on the rating
updates for all games played.

To illustrate, we consider a chess game played on June 1, 2022 on
\href{https://www.chess.com/}{Chess.com} by one of the authors, with
data obtained from the \pkg{chessR} package
(\protect\hyperlink{ref-R-chessR}{Zivkovic, 2022}) (see Section
\ref{sec:chess}). Prior to the game, the author was rated 1732, whereas
his opponent was rated 1683. Since both ratings are below 2400, we apply
a development coefficient of \(K=20\) to this example. The probability
of the author (\(a\)) defeating their opponent (\(b\)) was \[
\Pi_{ab} = \frac{1}{1 + 20^{(1683 - 1732)/400}} = 0.591 \,.
\] The author won the match: that outcome is associated with a score of
\(S_a = 1\). The post-game Elo rating for the author is thus \[
R^*_a = 1732 + 20(1 - 0.591) = 1740 \,.
\]

Besides chess, the Elo system has also been implemented to estimate team
strength in other sports. See Koning
(\protect\hyperlink{ref-koning2017rating}{2017}) for more information on
applications of the Elo rating in soccer, and Kovalchik \& Reid
(\protect\hyperlink{ref-kovalchik2019calibration}{2019}) and Kovalchik
(\protect\hyperlink{ref-kovalchik2016searching}{2016}) for Elo ratings
in tennis. Furthermore, Elo ratings are used extensively for rankings of
teams in numerous sports by the data journalists at
\href{https://fivethirtyeight.com}{FiveThirtyEight.com}.

\hypertarget{bayesian-state-space-models}{%
\subsection{Bayesian state-space
models}\label{bayesian-state-space-models}}

Glickman \& Stern (\protect\hyperlink{ref-Glickman1998}{1998}) propose a
Bayesian state-space model for paired comparisons for predicting NFL
games, allowing team strength parameters to vary over time. In
particular, they model point differential in the NFL by introducing
week-to-week and season-to-season as the two primary sources of
variation in team strengths. See also Glickman \& Stern
(\protect\hyperlink{ref-glickman2017estimating}{2017}) for more
discussion on estimating team strengths in American football.

More recently, Lopez et al. (\protect\hyperlink{ref-Lopez2018}{2018})
extend Glickman \& Stern (\protect\hyperlink{ref-Glickman1998}{1998})'s
state-space model to understand randomness in the four major American
sports leagues. Betting moneylines are used in place of point
differentials in order to estimate team strengths, and this framework
also accounts for home advantage. Both papers motivate the usefulness of
model-based measures of team strength by demonstrating their superiority
to low-resolution win-loss records. Apart from sports gambling, having
an accurate estimate of team strength is useful to team officials, who
are constantly monitoring and forecasting their team's ability.

In a similar Bayesian setting, Koopman \& Lit
(\protect\hyperlink{ref-koopman2015dynamic}{2015}) study English Premier
League soccer match results by assuming a bivariate Poisson distribution
with time-varying team abilities. This state-space approach appears to
improve on bookmaker's odds.

\hypertarget{sec:betting}{%
\section{Sports betting market data}\label{sec:betting}}

Most of the research in sports analytics is fueled by the analysis of
data recorded from the outcome of sports contests. However, a growing
body of literature is informed by data from sports betting markets.
Since the 2018 United States Supreme Court decision in
\href{https://en.wikipedia.org/wiki/Murphy_v._National_Collegiate_Athletic_Association}{\emph{Murphy
v. National Collegiate Athletic Association}}, sports gambling has
exploded in the U.S. The increasing interest in sports gambling has led
to increasing interest in sports gambling \emph{data}, and that data has
proven useful to researchers in at least two major ways.

First, betting market data is probably the best source for estimating
the true probability of a team winning a game. The efficiency of betting
market data in this respect has been demonstrated time and time again.
The utility of these estimates have then informed research that has
helped us learn about the sports themselves. In this sense, data
generated by sports gambling has been an important source of data useful
for sports analytics (see Section \ref{sec:betting-market-data}).

Second, sports analytics researchers have studied various types of
sports gambling outlets (including fantasy sports). This research has
estimated probabilities, evaluated common strategies, and offered
optimized approaches for a variety of different games of chance (see
Section \ref{sec:pools}). Some researchers have then tried to
demonstrate a positive return on some of these betting strategies, with
very limited success.

\hypertarget{sec:moneylines}{%
\subsection{Example: Win probabilities from betting market
data}\label{sec:moneylines}}

To see how betting market data can be used to estimate team strengths,
consider the betting lines posted on
\href{https://sportsbook.fanduel.com/navigation/nba?tab=nba-finals}{FanDuel
Sportsbook for the 2023 NBA Champion} on January 9, 2023 and shown in
Table \ref{tab:odds}. This is a futures market, because the actual NBA
champion will not be determined until June 2023. The Boston Celtics are
the favorite to win, with a moneyline of \(+390\), meaning that a \$100
bet on the Celtics to win the championship will pay back the original
bet and an additional \$390 if the Celtics win it all. This style of
odds are sometimes called
\href{https://en.wikipedia.org/wiki/Fixed-odds_betting\#Moneyline_odds}{\emph{American
odds}}. The corresponding fractional odds have the Celtics at 4.9:1 to
win the championship. Conversely, six teams share the lowest odds at
\(+50000\).

\begin{table}

\caption{\label{tab:odds}2023 NBA Championship odds for the top 6 and bottom 6 teams. Retrieved from FanDuel Sportsbook on January 9, 2023. }
\centering
\begin{tabular}[t]{llrrrr}
\toprule
Rank & Team & Line & Odds & Prob. & Prob. Normalized\\
\midrule
1 & Boston Celtics & 390 & 4.9 & 0.204 & 0.163\\
2 & Milwaukee Bucks & 500 & 6.0 & 0.167 & 0.133\\
3 & Brooklyn Nets & 800 & 9.0 & 0.111 & 0.089\\
4 & Golden State Warriors & 900 & 10.0 & 0.100 & 0.080\\
5 & Los Angeles Clippers & 1000 & 11.0 & 0.091 & 0.073\\
6 & Denver Nuggets & 1100 & 12.0 & 0.083 & 0.067\\
\addlinespace
25 & Oklahoma City Thunder & 50000 & 501.0 & 0.002 & 0.002\\
26 & Orlando Magic & 50000 & 501.0 & 0.002 & 0.002\\
27 & Charlotte Hornets & 50000 & 501.0 & 0.002 & 0.002\\
28 & Houston Rockets & 50000 & 501.0 & 0.002 & 0.002\\
29 & San Antonio Spurs & 50000 & 501.0 & 0.002 & 0.002\\
30 & Detroit Pistons & 50000 & 501.0 & 0.002 & 0.002\\
\addlinespace
Total & - & 496590 & 4995.9 & 1.253 & 1.000\\
\bottomrule
\end{tabular}
\end{table}

These moneylines (\(\ell_i\)) can be converted into an implied
probability (\(p_i\)) using the formula: \[
  p_i = \frac{100}{100 + \ell_i} \,.
\] The sum of those probabilities is greater than one---this is why the
sportsbook makes money regardless of who wins the championship. However,
the implied probabilities can be normalized by dividing by their sum to
recover true probabilities of each team winning the championship. Many
different researchers have shown that these normalized implied
probabilities are accurate, unbiased, and efficient estimates of the
true unknowable probabilities (see Lopez et al.
(\protect\hyperlink{ref-Lopez2018}{2018}) for discussion and an
extensive list of references).

In this case, the FanDuel futures market suggests that the Celtics have
a 16.3\% chance of winning the championship, while the Milwaukee Bucks
have the second best chance, at 13.3\%. These implied probabilities can
be used to fit various models for team strength, as described in Section
\ref{sec:strength}.

\hypertarget{sec:betting-market-data}{%
\subsection{The use of betting market data for sports
analytics}\label{sec:betting-market-data}}

While Lopez et al. (\protect\hyperlink{ref-Lopez2018}{2018}) use betting
market data to model team strengths, they do not directly address
strategies for betting or inefficiencies in betting markets. Early work
by Gandar et al. (\protect\hyperlink{ref-gandar1988testing}{1988})
examine the rationality of NFL betting markets and concludes that
statistical tests fail to reject the hypothesis of rationality. Related
work such as Lacey (\protect\hyperlink{ref-lacey1990estimation}{1990})
and Boulier et al. (\protect\hyperlink{ref-Boulier2006}{2006}) explores
the efficiency of NFL betting markets in the mid-1980s and late-1990s,
respectively. Neither paper finds strong evidence for inefficiencies in
the markets. Boulier \& Stekler
(\protect\hyperlink{ref-boulier2003predicting}{2003}) compare the
predictive performance of power rankings and media experts to the
betting market for NFL games and found that the betting market is the
best for predicting winners. Lopez \& Matthews
(\protect\hyperlink{ref-lopez2015building}{2015}) show that betting
market data was most useful in predicting men's college basketball
tournament outcomes.

Sports betting market data has also been used to investigate competitive
behavior within leagues. Soebbing \& Humphreys
(\protect\hyperlink{ref-soebbing2013gamblers}{2013}) find evidence that
sports bettors \emph{think} tanking in the NBA is occurring, although
the evidence for whether it actually is remains mixed.

\hypertarget{sec:pools}{%
\subsection{Analytics for sports betting}\label{sec:pools}}

Many different types of bets can be placed on sports. For individual
contests, bets involve point spreads, moneylines (see Section
\ref{sec:moneylines} for an example), odds, or other ways of
handicapping which team will win. Money can also be wagered on futures,
where odds are given in advance for events that may or may not transpire
(e.g., a certain team making the playoffs, or a certain player winning
the MVP award). Here, we focus on betting pools, in which a group of
people compete to predict winners in multiple contests (often a
tournament). We also address the inevitable question of whether
strategies exist that will consistently beat the market.

\hypertarget{betting-pools}{%
\subsubsection{Betting pools}\label{betting-pools}}

One popular type of betting pool is a survivor pool, in which
participants stay in the competition as long as they continue to
successfully pick winners. Bergman \& Imbrogno
(\protect\hyperlink{ref-Bergman2017SurvivingAN}{2017}) present formal
optimization approaches for NFL survivor pools and conclude that
planning for only part of the season yields optimal results in terms of
maximizing survival probability. Imbrogno \& Bergman
(\protect\hyperlink{ref-imbrogno2022computing}{2022}) estimate the
probability of having to share the winning pot in NFL survivor pools.

Perhaps the most commonly-studied sports betting market surrounds the
NCAA men's college basketball tournament. Breiter \& Carlin
(\protect\hyperlink{ref-breiter1997play}{1997}) use Monte Carlo methods
to study the standard ``office pool.'' Kaplan \& Garstka
(\protect\hyperlink{ref-kaplan2001march}{2001}) consider a variety of
NCAA college basketball pools, and find that the simple strategy of
picking the team with the better seed is generally, but not always,
optimal. Metrick (\protect\hyperlink{ref-metrick1996march}{1996}) finds
that bettors overback the heaviest favorites. Niemi et al.
(\protect\hyperlink{ref-niemi2008contrarian}{2008}) show an improved
return on investment by picking an undervalued champion and then
completing the rest of one's bracket by using published odds. Clair \&
Letscher (\protect\hyperlink{ref-clair2007optimal}{2007}) develop and
test strategies for maximizing expected return in both survivor and
tournament-style pools.

\hypertarget{beating-the-market}{%
\subsubsection{Beating the market}\label{beating-the-market}}

Naturally, after studying the efficiency of sports betting markets,
researchers try to find inefficiencies that can be exploited for
financial gain. Not surprisingly (given the efficiency of these
markets), such gains are hard to come by.

Sauer (\protect\hyperlink{ref-sauer1998economics}{1998}) finds that
while racetrack betting markets are generally efficient, information
asymmetry plays a role in creating inefficient markets. Nichols
(\protect\hyperlink{ref-nichols2014impact}{2014}) concludes that the
impact of travel is not completely incorporated into the betting
markets, but that any effect is too small to find any profitable
advantage. Paul \& Weinbach
(\protect\hyperlink{ref-PaulWeinbach2014}{2014}) investigate the
less-saturated betting market for the WNBA and fail to find strategies
for positive return on investment. Spann \& Skiera
(\protect\hyperlink{ref-spann2009sports}{2009}) show no way to beat the
market in the German premier soccer league, given the high fees
associated with placing bets.

More successfully, Buttrey (\protect\hyperlink{ref-Buttrey2016}{2016})
explores the NHL betting market and produces a model to predict win
probabilities in given games, then tests the model by placing market
price bets in games where the predicted probability differs from the
market. They find that their methods were able to produce a positive
return on investment.

\hypertarget{sec:tools}{%
\section{Tools}\label{sec:tools}}

Analytical work in sports requires facility with an ever-changing set of
computational tools for working with data. Sources of authoritative data
about sports are myriad, and are too numerous to list here. Software
tools for sports analytics are similarly numerous. For R, we maintain a
\href{https://CRAN.R-project.org/view=SportsAnalytics}{CRAN Task View
for Sports Analytics} that catalogs R packages published on the
Comprehensive R Archive Network (CRAN) and organizes them by sport
(\protect\hyperlink{ref-baumer2022ctv}{Baumer et al., 2022}). Table
\ref{tab:ctv} provides an overview of the currently available
sport-specific CRAN packages. Recently, Casals et al.
(\protect\hyperlink{ref-casals2022systematic}{2022}) offer a systematic
review of sport-related packages on CRAN. Further, a more general
collection of software tools is being curated by the SportsDataverse
initiative (\protect\hyperlink{ref-gilani2022sportsdataverse}{Gilani,
2022}).

\begin{table}

\caption{\label{tab:unnamed-chunk-7}A summary of sport-specific packages available on the Comprehensive R Archive Network (CRAN) as of October 16, 2022. While the major North American sports dominate the list, perhaps the fastest-growing collection is for esports. \label{tab:ctv}}
\centering
\begin{tabular}[t]{lr>{\raggedright\arraybackslash}p{7cm}}
\toprule
Sport & Number of Packages & List of Packages\\
\midrule
American Football & 12 & nflverse, nflfastR, nflreadr, nfl4th, nflseedR,
 nflplotR, NFLSimulatoR, fflr, ffscrapr, ffsimulator, gsisdecoder, cfbfastR\\
Association Football (Soccer) & 9 & worldfootballR, engsoccerdata, socceR, ggsoccer, footballpenaltiesBL, footBayes, itscalledsoccer, FPLdata, EUfootball\\
Basketball & 8 & BAwiR, AdvancedBasketballStats, uncmbb, BasketballAnalyzeR, NBAloveR, wehoop, hoopR, toRvik\\
Baseball/Softball & 7 & Lahman, retrosheet, pitchRx, mlbstats, baseballDBR, baseballr, runexp\\
Chess & 5 & chess, stockfish, bigchess, rchess, chessR\\
Esports & 5 & CSGo, rbedrock, ROpenData, opendotaR, RDota2\\
Hockey & 5 & hockeyR, NHLData, nhlapi, nhlscrape, fastRhockey\\
Cricket & 4 & yorkr, cricketr, cricketdata, howzatR\\
GPS Activity Tracking & 3 & trackeR, trackeRapp, rStrava\\
Track and Field & 2 & combinedevents, JumpeR\\
Australian Rules Football & 1 & fitzRoy\\
Swimming & 1 & SwimmeR\\
Volleyball & 1 & volleystat\\
\bottomrule
\end{tabular}
\end{table}

In the remainder of this section, we highlight a few tools for sports
analytics that are of general interest and illustrate a common paradigm
for how these tools can be used in conjunction.

\hypertarget{sec:chess}{%
\subsection{Case study in how tools fit together:
chess}\label{sec:chess}}

Many tools in sports analytics provide the ability to read, write, and
plot data stored in a sport-specific format. For example, consider
chess, where the sequence of moves in games is often recorded in
\href{https://en.wikipedia.org/wiki/Portable_Game_Notation}{Portable
Game Notation} (PGN). Software tools can then be built around this
well-defined format. The \pkg{chess} package
(\protect\hyperlink{ref-R-chess}{Lente, 2020}) provides R users with the
ability to read, write, display, and manipulate chess data in PGN
format.

\href{https://en.wikipedia.org/wiki/API}{Application programming
interfaces} (APIs) are also a common source for data retrieval. In
chess, the \pkg{chessR} package
(\protect\hyperlink{ref-R-chessR}{Zivkovic, 2022}) allows R users to
download game data from the Chess.com API. This type of infrastructure,
where one package is the ``workhorse'' that facilitates common generic
data operations, and other packages layer on specific functionality, is
common in sports analytics. See Section \ref{sec:elo} for an example of
how the \pkg{chessR} package can be used to compute Elo ratings.

\begin{figure}

{\centering \includegraphics[width=0.49\linewidth]{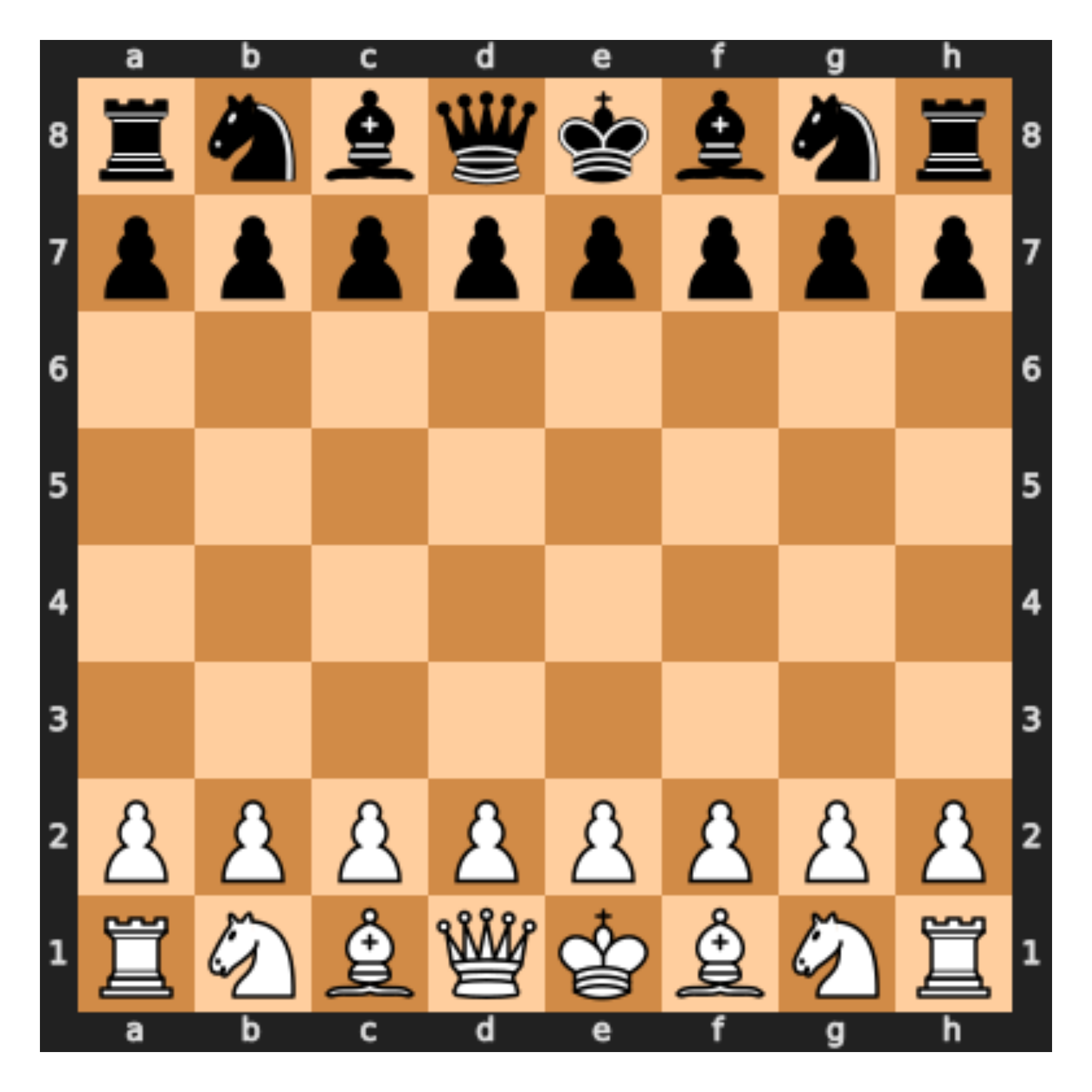} \includegraphics[width=0.49\linewidth]{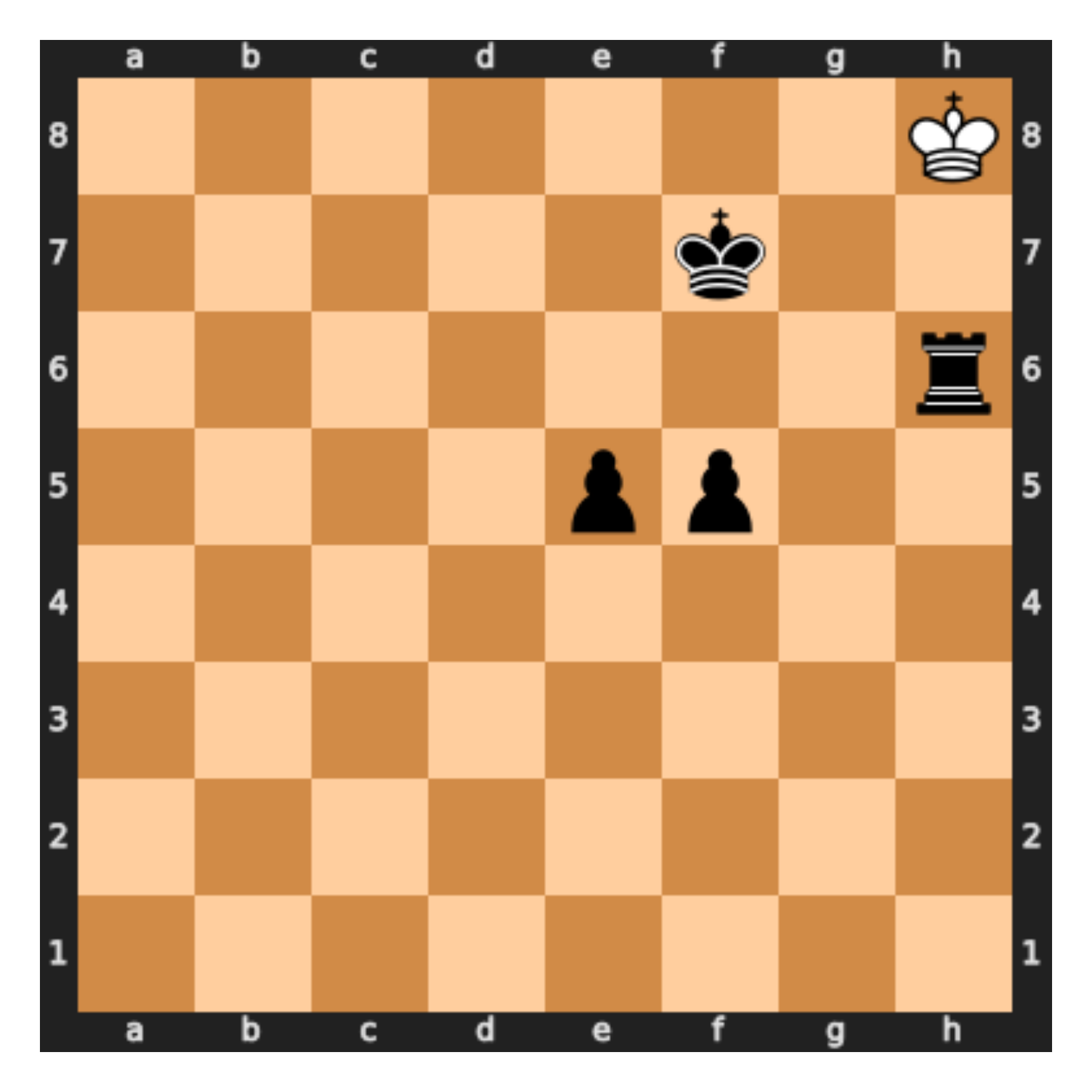} 

}

\caption{At left, the starting chess board printed via the \pkg{chess} package. At right, the final position for one of the authors' recent wins (a checkmate playing Black).}\label{fig:chess}
\end{figure}

Figure \ref{fig:chess} shows a rendering of the starting chess board
obtained via the \pkg{chess} package, along with the final position in
the game won by one of the authors mentioned earlier in Section
\ref{sec:elo} (with data downloaded via the \pkg{chessR} package). We
note how the contextual information provided by the chessboard is
instrumental in helping the reader understand the data (How many of us
can visualize PGN directly?). In Section \ref{sec:graphics}, we outline
a collection of graphical tools that provide similar context for
different playing surfaces.

\hypertarget{sec:graphics}{%
\subsection{Graphical tools}\label{sec:graphics}}

Creating effective data graphics is a key component of statistical
communication, and sports is no exception. We highlight a few packages
that assist with the creation of data graphics about sports.

\begin{figure}

{\centering \includegraphics{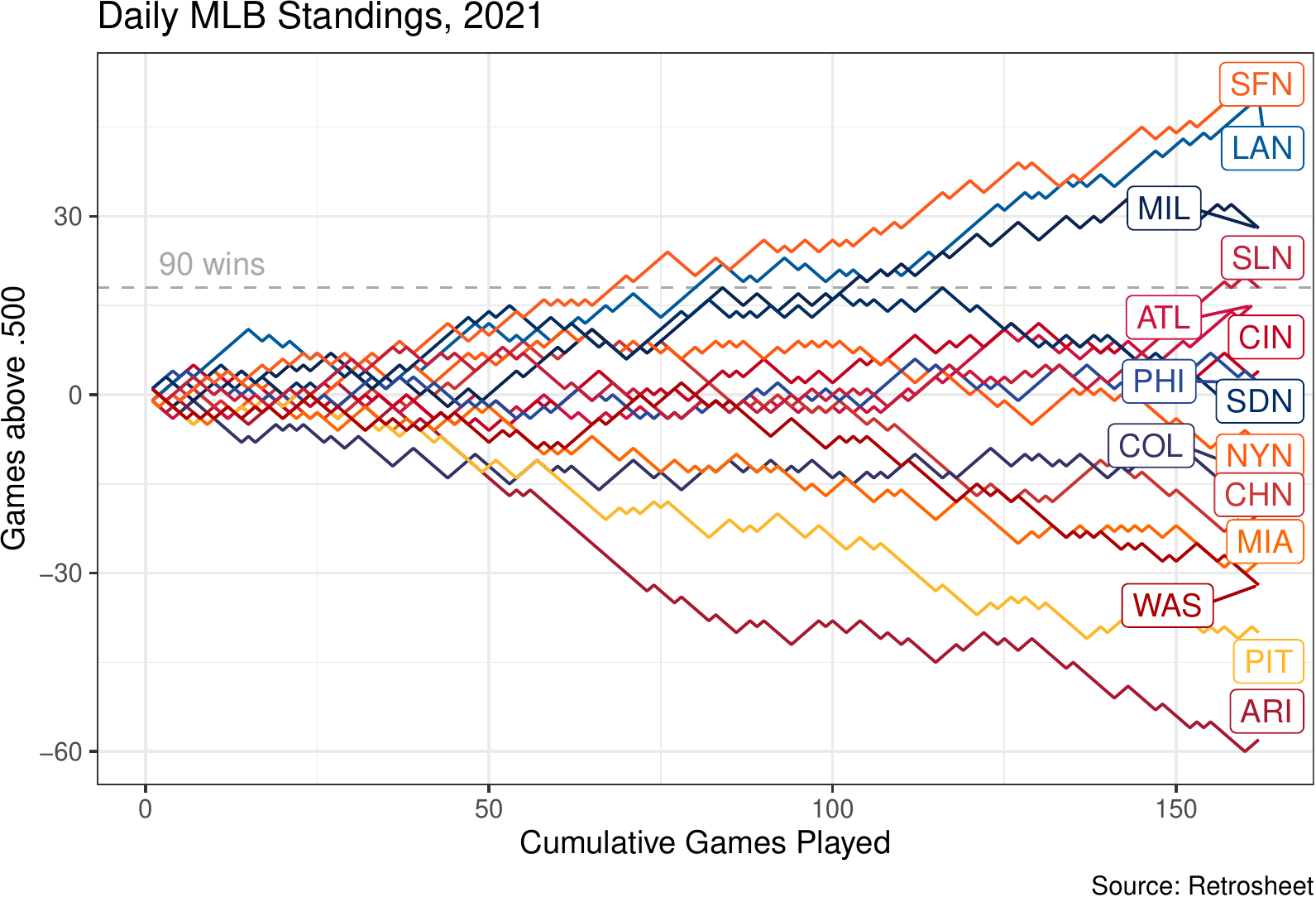} 

}

\caption{The progression of National Leauge team standings during the 2021 Major League Baseball season. Note how the use of team colors makes it possible to untangle what would otherwise be a messy jumble of indistinguishable lines. Data provided by \pkg{retrosheet} and colors provided by \pkg{teamcolors}.}\label{fig:mlb_standings}
\end{figure}

Each professional sports team has its own brand, most obviously
identified by a team logo and set of colors. The \pkg{teamcolors} R
package (\protect\hyperlink{ref-R-teamcolors}{Baumer \& Matthews, 2020})
provides color palettes and logos for men's and women's professional and
collegiate sports teams, as well as color and fill scale functions
compatible with \pkg{ggplot2} (\protect\hyperlink{ref-R-ggplot2}{Wickham
et al., 2022}). For example, the NFL teams' colors and logos shown in
Figure \ref{fig:nflfastr-wp} were provided by the \pkg{teamcolors}
package. Figure \ref{fig:mlb_standings} illustrates how the use of team
colors, which have a natural association for many sports fans, can help
to untangle what would otherwise be messy data graphics. In Figure
\ref{fig:mlb_standings}, 30 different lines are plotted on top of one
another, crisscrossing and intersecting in various unpredictable ways.
However, the use of team colors to identify the lines makes it possible
to follow the trajectory of most teams over the course of the season.

\begin{figure}

{\centering \includegraphics{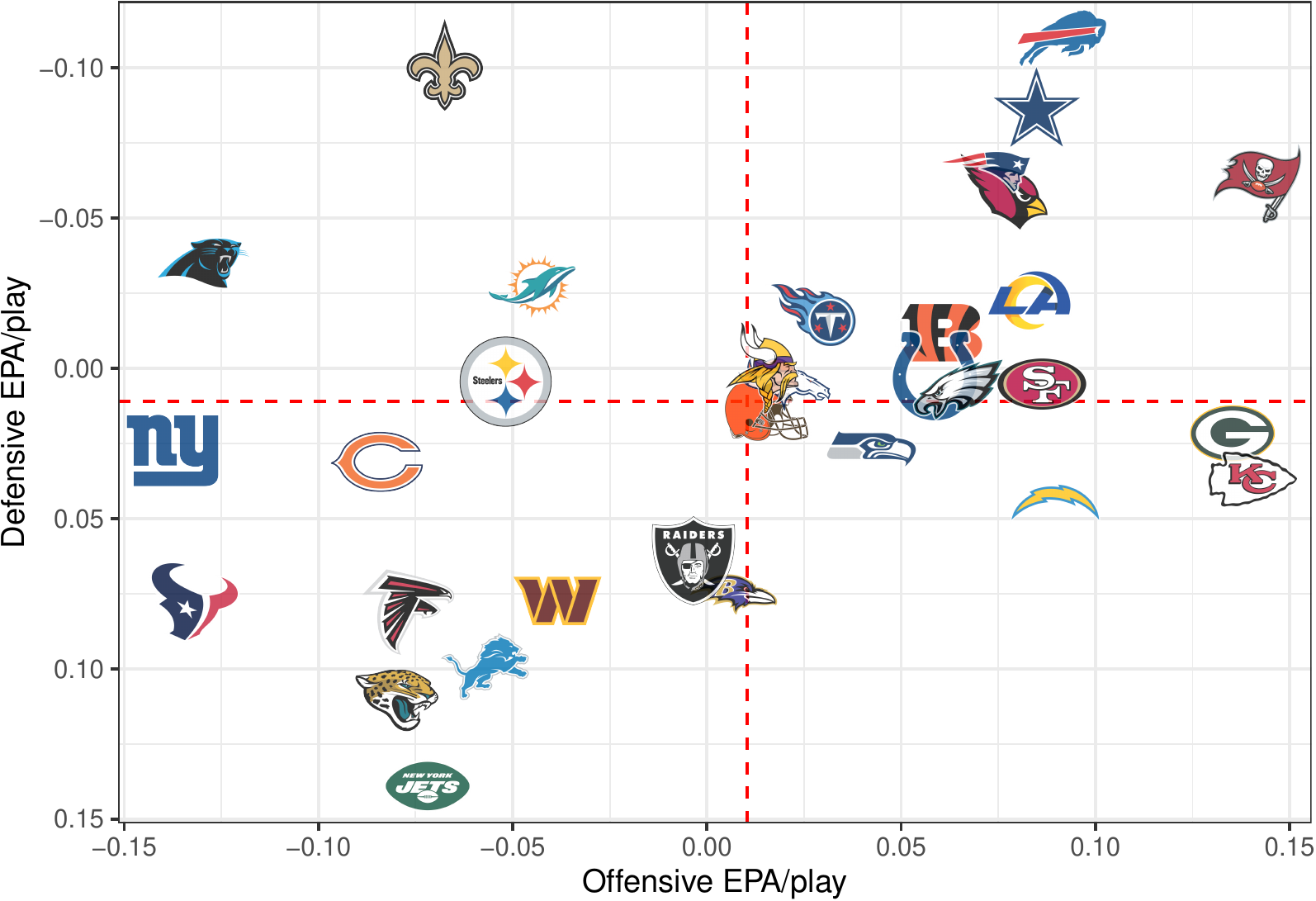} 

}

\caption{Offensive and defensive expected points added per play for the 2021 NFL regular season, plotted with \pkg{nflplotR} using data from \pkg{nflfastR}.}\label{fig:nflplotr}
\end{figure}

\pkg{nflplotR} (\protect\hyperlink{ref-R-nflplotR}{Carl, 2022}) has a
similar goal to \pkg{teamcolors}. It also provides \pkg{ggplot2}
extensions but is designed specifically for the NFL. A great feature of
\pkg{nflplotR} is the collection of \texttt{geom\_*()} (geometric
object) functions that enhance high-quality plotting of NFL team logos
and player images with \pkg{ggplot2}. Figure \ref{fig:nflplotr} shows a
scatterplot of offensive and defensive expected points added for NFL
teams in the 2021 regular season. The logos of all 32 American football
clubs are plotted in place of the usual dots, making it easier for the
reader to identify which team each data point represents.

\begin{figure}

{\centering \includegraphics[width=0.49\linewidth]{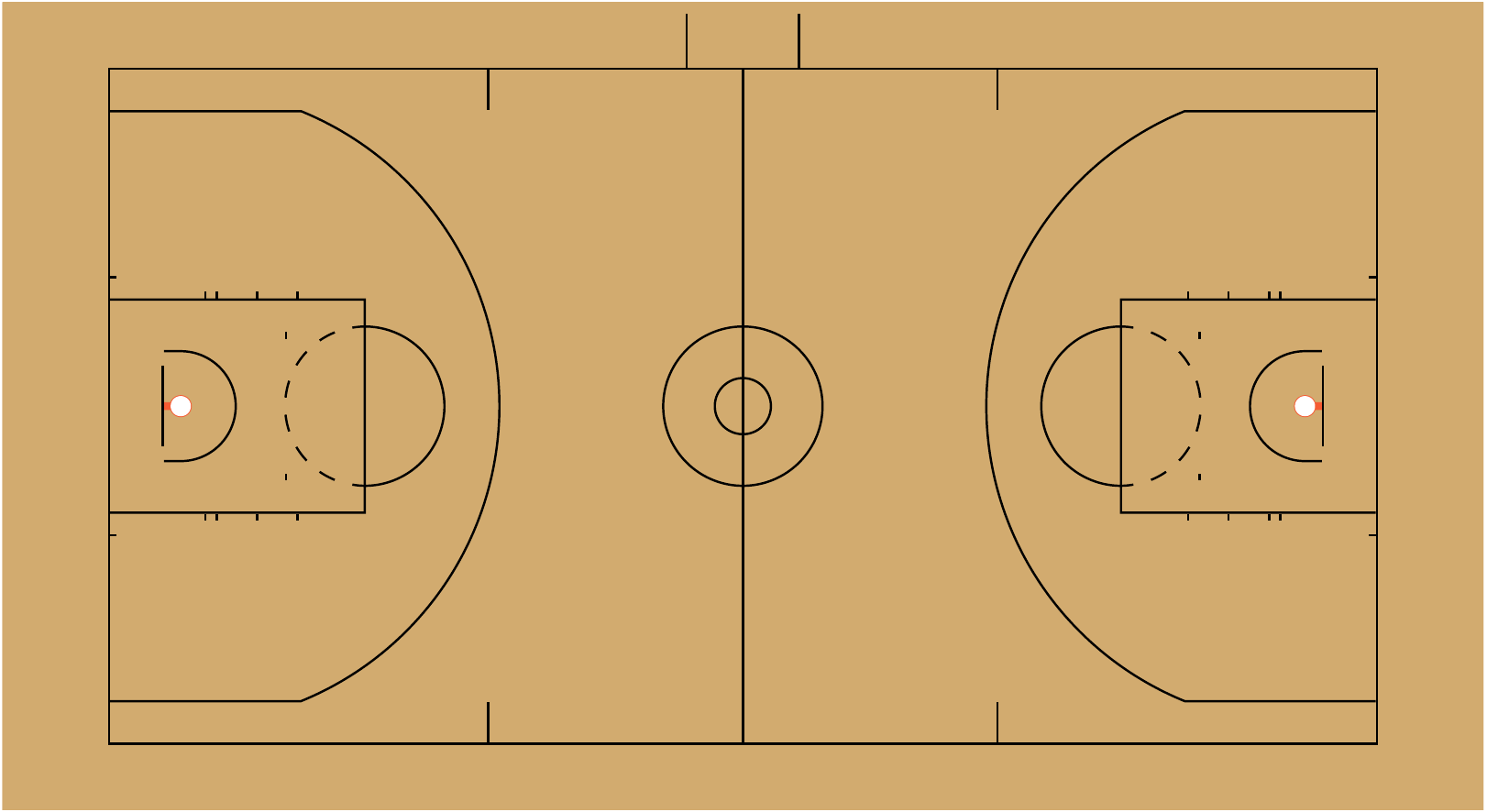} \includegraphics[width=0.49\linewidth]{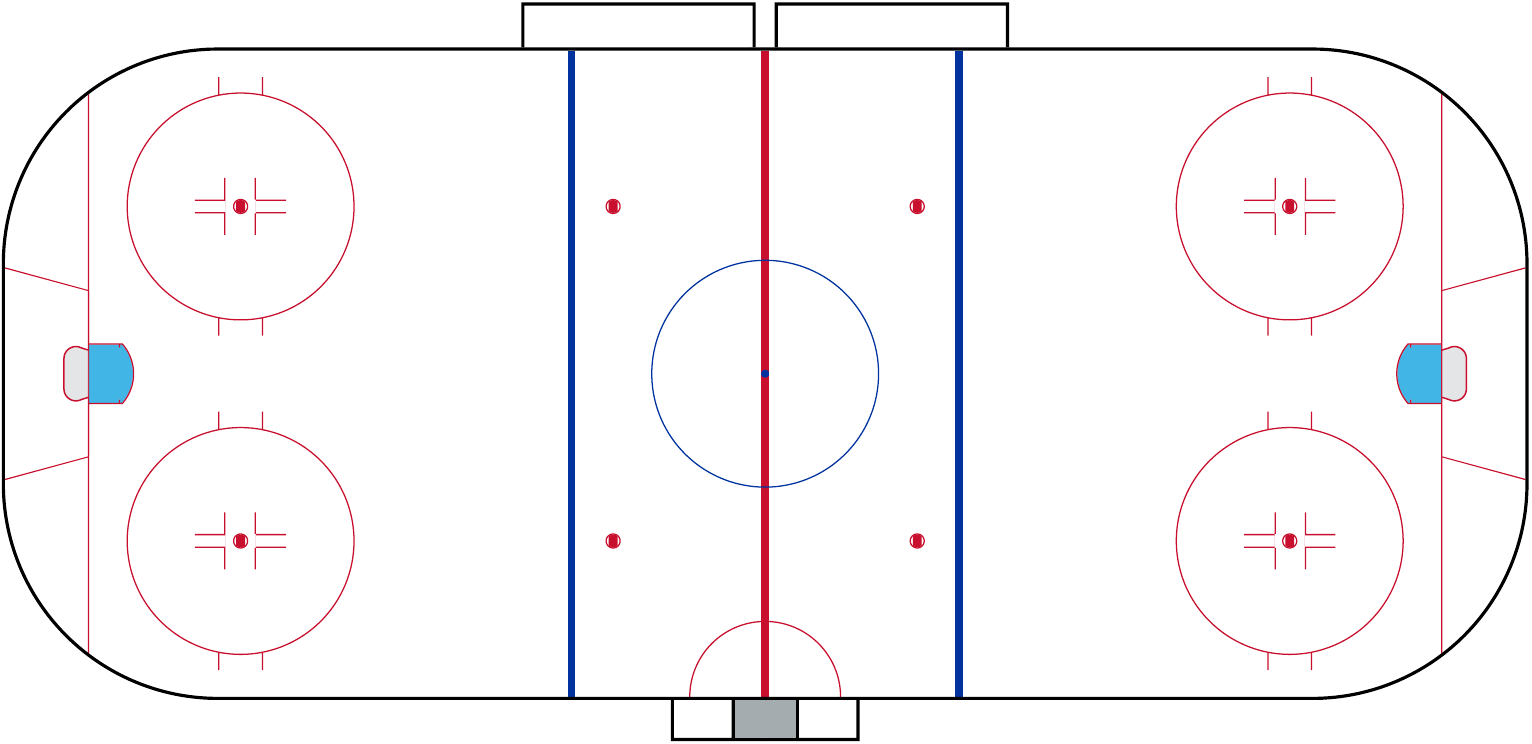} 

}

\caption{At left, an NBA basketball court drawn by \pkg{sportyR}. At right, an NHL hockey rink drawn by \pkg{sportyR}.}\label{fig:sportyR}
\end{figure}

Player tracking data contains coordinates that reveal player movement,
and these coordinates are always understood in context relative to
reference points on the field, court, ice, board, or pitch for a
particular sport. Orienting these points graphically may require drawing
a complex set of guidelines that provide that context to readers.
Thankfully, the \pkg{sportyR} package
(\protect\hyperlink{ref-R-sportyR}{Drucker, 2022}) contains generic
playing surfaces for baseball, basketball, curling, American football,
ice hockey, soccer, and tennis that can be added to \pkg{ggplot}
graphics with a single function call. The surfaces plotted in Figure
\ref{fig:sportyR} are helpful in contextualizing player tracking data
(such as those shown in Figure \ref{fig:epv-demo}) and would be
laborious for each analyst to have to create on their own. With the
increased availability of player tracking data, this particular tool
should see increased usage.

\hypertarget{sec:baseball}{%
\subsection{Case study in the evolution of tools and research:
baseball}\label{sec:baseball}}

As the granularity of baseball data has evolved over time, so too have
the statistical methodologies for modeling that data, and the tools for
working with it.

For example, before George Lindsey's work (see Section \ref{sec:ev}),
most of the baseball data that was publicly available was seasonal: it
showed only season totals for each player. These data, now available
through the \pkg{Lahman} package
(\protect\hyperlink{ref-R-Lahman}{Friendly et al., 2022}), were
sufficient to study broad trends in baseball, and led to insights such
as the value of expected winning percentage (see Section \ref{sec:wpct})
and the importance of on-base percentage relative to batting average.
These relatively simple insights fueled the ``Moneyball''
(\protect\hyperlink{ref-lewis2004moneyball}{Lewis, 2004}) era revolution
in sports analytics (\protect\hyperlink{ref-baumer2014sabermetric}{B.
Baumer \& Zimbalist, 2014}).

Over time, the resolution of baseball data has improved to include
play-by-play data, pitch-by-pitch data, and now player tracking data.

The \pkg{retrosheet} package
(\protect\hyperlink{ref-R-retrosheet}{Douglas \& Scriven, 2021}) now
provides access to the historical play-by-play data available from
Retrosheet (this is a comprehensive version of what Lindsay collected
for his research). This play-by-play data allowed researchers to learn
about strategies, like those that we discussed in Section \ref{sec:ev}.
In baseball, this deepened our understanding of bunting, stolen bases,
handedness, batting order, and many other aspects of the game.
Play-by-play data underlies much of the analysis in Tango et al.
(\protect\hyperlink{ref-Tango2007book}{2007}).

The \pkg{pitchRx} package (\protect\hyperlink{ref-R-pitchRx}{Sievert,
2015}) provides access to pitch-by-pitch data that fueled innovative
research into catcher framing
(\protect\hyperlink{ref-deshpande2017hierarchical}{Deshpande \& Wyner,
2017}), pitch values (\protect\hyperlink{ref-healey2019bayesian}{Healey,
2019}), and pitch classification
(\protect\hyperlink{ref-sidle2018using}{Sidle \& Tran, 2018}). Catcher
framing is a notable example of a concept that scouts talked about for
decades, but that could not be quantified by analysts until data of the
appropriate resolution became available.

While play-by-play data allows us to make valuations \emph{between
plays}, player tracking data allows us to make valuations \emph{within
plays}. The \pkg{baseballr} package
(\protect\hyperlink{ref-R-baseballr}{Petti \& Gilani, 2022}) now
provides access to player tracking data from Statcast. These data have
led to investigations into how defensive shifts affect batting
performance (\protect\hyperlink{ref-bouzarth2021swing}{Bouzarth et al.,
2021}), as well as how launch angles affect the probability of hitting a
home run (\protect\hyperlink{ref-marchi2018analyzing}{Marchi et al.,
2018}).

As we saw above with chess, the packages in baseball fit together in
creative ways. In Figure \ref{fig:mlb_standings}, we showed how
\pkg{teamcolors} can illuminate data pulled from \pkg{retrosheet} to
make an informative data graphic. One could just as easily use
\pkg{sportyR} to generate a field graphic, and then overlay player
tracking data obtained from \pkg{baseballr} to depict defensive shifts.

Thus, these R packages enable research by making data more easily
available. Moreover, because R is scriptable, they make it easier to
share research that is reproducible. Recent conferences, such as the
\href{https://www.stat.cmu.edu/cmsac/conference/2021/}{Carnegie Mellon
Sports Analytics Conference}, have included a reproducible research
competition to foster these efforts (see Section
\ref{sec:opportunities}).

\hypertarget{sec:opportunities}{%
\section{Opportunities}\label{sec:opportunities}}

Public research in sports analytics is driven in part by several notable
competitions and conferences. These venues have been an important source
of new ideas and have contributed to the diversification of the field by
breaking down barriers to entry.

In 2014, Kaggle launched its first March Machine Learning Mania
competition for predicting the outcome of the NCAA men's basketball
tournament. 243 entrants competed for the \$15,000 cash prize by
submitting predicted probabilities for every possible pairwise matchup
among the 68 college basketball teams in the tournament
(\protect\hyperlink{ref-GlickmanSonas2015}{Glickman \& Sonas, 2015}).
Subsequently, the \emph{Journal of Quantitative Analysis in Sports}
(JQAS) released a
\href{https://www.degruyter.com/journal/key/jqas/11/1/html}{special
issue on prediction methodology for the NCAA men's basketball
tournament}. Among the published papers, we learned that the winning
entry was based on a fairly simple logistic regression model trained on
betting market data (\protect\hyperlink{ref-lopez2015building}{Lopez \&
Matthews, 2015}). Thus, the competition not only sparked interest in
sports analytics, but also resulted in peer-reviewed research which, in
that case, demonstrated the value of quality data over sophisticated
modeling.

Perhaps motivated by his success in the Kaggle March Madness
competition, Michael Lopez joined the National Football League and
launched the
\href{https://operations.nfl.com/gameday/analytics/big-data-bowl/}{Big
Data Bowl} in 2019. This annual competition has similarly fueled new
research directions in American football and a
\href{https://www.degruyter.com/journal/key/jqas/16/2/html}{\emph{JQAS}
special issue on player tracking data in the National Football League}
(\protect\hyperlink{ref-lopez2020bigger}{Lopez, 2020}). Successful
entries and their corresponding publications
(\protect\hyperlink{ref-ChuReyersThomsonWu2020}{Chu et al., 2020};
\protect\hyperlink{ref-DeshpandeEvans2020}{Deshpande \& Evans, 2020};
\protect\hyperlink{ref-Yurko2020going}{Yurko et al., 2020}) have
launched the careers of several of the most prominent early-career
researchers in sports analytics.

Similar competitions that provide opportunities for aspiring researchers
to tackle sports analytics problems include the
\href{https://www.stathletes.com/big-data-cup/}{Big Data Cup} for ice
hockey and the \href{https://sabr.org/analytics/case}{SABR Diamond
Dollars Case Competition} for baseball, and the
\href{https://www.kaggle.com/competitions/big-data-derby-2022}{Big Data
Derby} for horse racing.

As the field of sports analytics has grown, a proliferation of sports
specific and regional sports analytics conferences have arisen. The
biennial \href{https://www.nessis.org/}{New England Symposium on
Statistics in Sports} is likely the longest-running academic conference
devoted to sports analytics. Its West Coast counterpart is
\href{http://www.cascadiasports.com/}{The Cascadia Symposium on
Statistics in Sports}. Many influential results have been showcased for
the first time at these conferences. Other prominent sports analytics
conferences include the
\href{https://www.stat.cmu.edu/cmsac/conference/2022/}{Carnegie Mellon
Sports Analytics Conference},
\href{https://statds.org/events/ucsas2022/}{UConn Sports Analytics
Symposium}, and \href{http://www.mathsportinternational.com/}{MathSport
International}.

The highest-profile sports analytics conference is undoubtedly the
\href{https://www.sloansportsconference.com/}{Sloan Sports Analytics
Conference}, which draws academics, industry professionals, vendors, and
media organizations. While the conference holds a research competition
and has certainly drawn attention to sports analytics, it has also been
criticized for a variety of shortcomings. These criticisms include a
lack of emphasis on reproducibility in the research competition, high
ticket prices, the large salaries taken by the organizers, and the lack
of diversity among attendees and presenters
(\protect\hyperlink{ref-funt2022}{Funt, 2022}).

It is also worth noting that a significant, but unknown, proportion of
the most innovative research is being conducted by professional sports
teams. This research will likely never be published, because each team
will use it to their competitive advantage. Part of what enables this
research is better data. For example, professional sports teams can
collect biometric data on their own players, and use that data to learn
about how their workouts, sleep patterns, and diets impact their
athletic performance. While this research may constitute ``emerging
methodologies,'' it unfortunately will take years, if at all, before the
public benefits from it.

\hypertarget{sec:conclusion}{%
\section{Conclusion}\label{sec:conclusion}}

As an applied science, sports analytics may lack a
\href{https://en.wikipedia.org/wiki/Grand_Unified_Theory}{grand unified
theory} that succinctly characterizes game play across all sports.
However, as a maturing discipline, sports analytics has been able to
address fundamental questions common to many sports. In this paper, we
explore three of those big questions: Who are the best teams and how
good are they? What is the likelihood of each team winning the game at
any given juncture? Is there a generic framework for evaluating
strategies at any given juncture in a game?

Other fundamental questions are addressed elsewhere. How significant is
the element of chance in a particular sport? Given that we know who the
best teams are, who are the best players and how can we quantify their
relative contributions? What combinations of players work best together
in a particular sport? In particular, see Lopez et al.
(\protect\hyperlink{ref-Lopez2018}{2018}) for estimations of the element
of chance across four major sports. The second question is often
addressed using a formulation of \emph{wins above replacement}
(WAR)---see Baumer et al.
(\protect\hyperlink{ref-baumer2015openwar}{2015}) and Yurko et al.
(\protect\hyperlink{ref-Yurko2019nflwar}{2019}) for details in baseball
and American football. Recent work by Che \& Glickman
(\protect\hyperlink{ref-che2022athlete}{2022}) also addresses this
question across sports. The third question is most compelling in sports
like basketball, ice hockey, and soccer, where substitutions are common
and it is obvious that different combinations of players with different
sets of skills will result in squad of varying strengths and weaknesses.
The concept of \emph{plus-minus}, and then \emph{adjusted} plus-minus is
frequently applied to address this question (see Hvattum
(\protect\hyperlink{ref-hvattum2019comprehensive}{2019}) for a
comprehensive overview of applications).

In drawing together these three big ideas in sports analytics, we have
also drawn attention to new uses of sports betting market data, some
computational tools for doing sports analytics work, and opportunities
to showcase that work. Our discussion in Section \ref{sec:baseball}
shows how the increased resolution of available data has catalyzed new
research directions in baseball, but this same dynamic is playing out in
all sports. It is through these exchanges of ideas, tools, models, and
data that analytics moves our collective understanding of sports
forward.

\hypertarget{acknowledgments}{%
\section*{Acknowledgments}\label{acknowledgments}}
\addcontentsline{toc}{section}{Acknowledgments}

We are grateful to Michael Lopez and Katherine Evans for their thoughts
on early versions of this paper.

The R Markdown file that generated this paper, including all R code, is
available at \url{https://github.com/beanumber/wire21}.

\hypertarget{references}{%
\section*{References}\label{references}}
\addcontentsline{toc}{section}{References}

\hypertarget{refs}{}
\begin{CSLReferences}{1}{0}
\leavevmode\vadjust pre{\hypertarget{ref-agresti2018introduction}{}}%
Agresti, A. (2018). \emph{An introduction to categorical data analysis}
(3rd ed.). Hoboken, NJ: John Wiley \& Sons.

\leavevmode\vadjust pre{\hypertarget{ref-Alamar2010}{}}%
Alamar, B. (2010). Measuring risk in NFL playcalling. \emph{Journal of
Quantitative Analysis in Sports}, \emph{6}(2), 11.
\url{https://doi.org/10.2202/1559-0410.1235}

\leavevmode\vadjust pre{\hypertarget{ref-Albert2015}{}}%
Albert, J. (2015). Player evaluation using win probabilities in sports
competitions. \emph{Wiley Interdisciplinary Reviews: Computational
Statistics}, \emph{7}(5), 316--325.
\url{https://doi.org/10.1002/wics.1358}

\leavevmode\vadjust pre{\hypertarget{ref-Albert2001curve}{}}%
Albert, J., \& Bennett, J. (2001). \emph{Curve ball: Baseball,
statistics, and the role of chance in the game}. New York: Springer.

\leavevmode\vadjust pre{\hypertarget{ref-Albert2017handbook}{}}%
Albert, J., Glickman, M. E., Swartz, T. B., \& Koning, R. H. (2016).
\emph{Handbook of statistical methods and analyses in sports} (p. 520).
New York: Chapman; Hall/CRC Press.
\url{https://doi.org/10.1201/9781315166070}

\leavevmode\vadjust pre{\hypertarget{ref-baumer2015openwar}{}}%
Baumer, B. S., Jensen, S. T., \& Matthews, G. J. (2015). {openWAR}: An
open source system for evaluating overall player performance in {M}ajor
{L}eague {B}aseball. \emph{Journal of Quantitative Analysis in Sports},
\emph{11}(2), 69--84. \url{https://doi.org/10.1515/jqas-2014-0098}

\leavevmode\vadjust pre{\hypertarget{ref-baumer2021mdsr}{}}%
Baumer, B. S., Kaplan, D. T., \& Horton, N. J. (2021). \emph{{Modern
Data Science with R}} (2nd ed., pp. 1--673). Boca Raton, FL: Chapman;
Hall/CRC Press. \url{https://mdsr-book.github.io/mdsr2e/}

\leavevmode\vadjust pre{\hypertarget{ref-R-teamcolors}{}}%
Baumer, B. S., \& Matthews, G. J. (2020). \emph{Teamcolors: Color
palettes for pro sports teams}.
\url{http://github.com/beanumber/teamcolors}

\leavevmode\vadjust pre{\hypertarget{ref-baumer2022ctv}{}}%
Baumer, B. S., Nguyen, Q., \& Matthews, G. J. (2022). \emph{CRAN task
view: Sports analytics}.
\url{https://CRAN.R-project.org/view=SportsAnalytics}

\leavevmode\vadjust pre{\hypertarget{ref-baumer2014sabermetric}{}}%
Baumer, B., \& Zimbalist, A. (2014). \emph{The sabermetric revolution:
Assessing the growth of analytics in baseball} (p. 187). Philadelphia:
University of Pennsylvania Press.

\leavevmode\vadjust pre{\hypertarget{ref-beaudoin2010strategies}{}}%
Beaudoin, D., \& Swartz, T. B. (2010). Strategies for pulling the goalie
in hockey. \emph{The American Statistician}, \emph{64}(3), 197--204.
\url{https://doi.org/10.1198/tast.2010.09147}

\leavevmode\vadjust pre{\hypertarget{ref-Bergman2017SurvivingAN}{}}%
Bergman, D., \& Imbrogno, J. (2017). Surviving a national football
league survivor pool. \emph{Oper. Res.}, \emph{65}, 1343--1354.
\url{https://doi.org/10.1287/opre.2017.1633}

\leavevmode\vadjust pre{\hypertarget{ref-bornn2017studying}{}}%
Bornn, L., Cervone, D., Franks, A., \& Miller, A. (2017). Studying
basketball through the lens of player tracking data. In \emph{Handbook
of statistical methods and analyses in sports} (pp. 261--286). Boca
Raton, FL: CRC Press.

\leavevmode\vadjust pre{\hypertarget{ref-boulier2003predicting}{}}%
Boulier, B. L., \& Stekler, H. O. (2003). Predicting the outcomes of
national football league games. \emph{International Journal of
Forecasting}, \emph{19}(2), 257--270.
\url{https://doi.org/10.1016/s0169-2070(01)00144-3}

\leavevmode\vadjust pre{\hypertarget{ref-Boulier2006}{}}%
Boulier, B. L., Stekler, H. O., \& Amundson, S. (2006). Testing the
efficiency of the national football league betting market. \emph{Applied
Economics}, \emph{38}(3), 279--284.
\url{https://doi.org/10.1080/00036840500368904}

\leavevmode\vadjust pre{\hypertarget{ref-bouzarth2021swing}{}}%
Bouzarth, E., Grannan, B., Harris, J., Hartley, A., Hutson, K., \&
Morton, E. (2021). Swing shift: A mathematical approach to defensive
positioning in baseball. \emph{Journal of Quantitative Analysis in
Sports}, \emph{17}(1), 47--55.
\url{https://doi.org/10.1515/jqas-2020-0027}

\leavevmode\vadjust pre{\hypertarget{ref-Bradley1952}{}}%
Bradley, R. A., \& Terry, M. E. (1952). Rank analysis of incomplete
block designs: I. The method of paired comparisons. \emph{Biometrika},
\emph{39}(3/4), 324--345. \url{https://doi.org/10.2307/2334029}

\leavevmode\vadjust pre{\hypertarget{ref-breiter1997play}{}}%
Breiter, D. J., \& Carlin, B. P. (1997). How to play office pools if you
must. \emph{Chance}, \emph{10}(1), 5--11.
\url{https://doi.org/10.1080/09332480.1997.10554789}

\leavevmode\vadjust pre{\hypertarget{ref-Brenzel2019}{}}%
Brenzel, P., Shock, W., \& Yang, H. (2019). An analysis of curling using
a three-dimensional markov model. \emph{Journal of Sports Analytics},
\emph{5}(2), 101--119. \url{https://doi.org/10.3233/jsa-180279}

\leavevmode\vadjust pre{\hypertarget{ref-Buttrey2016}{}}%
Buttrey, S. E. (2016). Beating the market betting on {NHL} hockey games.
\emph{Journal of Quantitative Analysis in Sports}, \emph{12}(2), 87--98.
\url{https://doi.org/10.1515/jqas-2015-0003}

\leavevmode\vadjust pre{\hypertarget{ref-R-nflplotR}{}}%
Carl, S. (2022). \emph{nflplotR: NFL logo plots in ggplot2}.
\url{https://CRAN.R-project.org/package=nflplotR}

\leavevmode\vadjust pre{\hypertarget{ref-R-nflfastR}{}}%
Carl, S., \& Baldwin, B. (2022). \emph{nflfastR: Functions to
efficiently access NFL play by play data}.
\url{https://CRAN.R-project.org/package=nflfastR}

\leavevmode\vadjust pre{\hypertarget{ref-caro2013testing}{}}%
Caro, C. A., Machtmes, R., et al. (2013). Testing the utility of the
pythagorean expectation formula on division one college football: An
examination and comparison to the morey model. \emph{Journal of Business
\& Economics Research (JBER)}, \emph{11}(12), 537--542.
\url{https://doi.org/10.19030/jber.v11i12.8261}

\leavevmode\vadjust pre{\hypertarget{ref-Carroll1988hidden}{}}%
Carroll, B. N., Palmer, P., Thorn, J., \& Pietrusza, D. (1988).
\emph{\href{}{The hidden game of football}} (p. 415). New York: Total
Sports, Inc.

\leavevmode\vadjust pre{\hypertarget{ref-Carter1971operations}{}}%
Carter, Virgil., \& Machol, R. E. (1971). Technical note--operations
research on football. \emph{Operations Research}, \emph{19}(2),
541--544. \url{https://doi.org/10.1287/opre.19.2.541}

\leavevmode\vadjust pre{\hypertarget{ref-casals2022systematic}{}}%
Casals, M., Fernández, J., Martínez, V., Lopez, M., Langohr, K., \&
Cortés, J. (2022). A systematic review of sport-related packages within
the {R CRAN} repository. \emph{International Journal of Sports Science
\& Coaching}, 1. \url{https://doi.org/10.1177/17479541221136238}

\leavevmode\vadjust pre{\hypertarget{ref-cervone2014pointwise}{}}%
Cervone, D., D'Amour, A., Bornn, L., \& Goldsberry, K. (2014).
\emph{Pointwise: Predicting points and valuing decisions in real time
with {NBA} optical tracking data}. \emph{28}, 3.
\url{http://www.lukebornn.com/papers/cervone_ssac_2014.pdf}

\leavevmode\vadjust pre{\hypertarget{ref-cervone2016multiresolution}{}}%
Cervone, D., D'Amour, A., Bornn, L., \& Goldsberry, K. (2016). A
multiresolution stochastic process model for predicting basketball
possession outcomes. \emph{Journal of the American Statistical
Association}, \emph{111}(514), 585--599.
\url{https://doi.org/10.1080/01621459.2016.1141685}

\leavevmode\vadjust pre{\hypertarget{ref-che2022athlete}{}}%
Che, J., \& Glickman, M. (2022). Athlete rating in multi-competitor
games with scored outcomes via monotone transformations. \emph{arXiv
Preprint arXiv:2205.10746}. \url{https://arxiv.org/pdf/2205.10746}

\leavevmode\vadjust pre{\hypertarget{ref-R-xgboost}{}}%
Chen, T., He, T., Benesty, M., Khotilovich, V., Tang, Y., Cho, H., Chen,
K., Mitchell, R., Cano, I., Zhou, T., Li, M., Xie, J., Lin, M., Geng,
Y., Li, Y., \& Yuan, J. (2022). \emph{Xgboost: Extreme gradient
boosting}. \url{https://github.com/dmlc/xgboost}

\leavevmode\vadjust pre{\hypertarget{ref-ChuReyersThomsonWu2020}{}}%
Chu, D., Reyers, M., Thomson, J., \& Wu, L. Y. (2020). Route
identification in the {National Football League}: An application of
model-based curve clustering using the EM algorithm. \emph{Journal of
Quantitative Analysis in Sports}, \emph{16}(2), 121--132.
\url{https://doi.org/10.1515/jqas-2019-0047}

\leavevmode\vadjust pre{\hypertarget{ref-clair2007optimal}{}}%
Clair, B., \& Letscher, D. (2007). Optimal strategies for sports betting
pools. \emph{Operations Research}, \emph{55}(6), 1163--1177.
\url{https://doi.org/10.1287/opre.1070.0448}

\leavevmode\vadjust pre{\hypertarget{ref-clark2020bayesian}{}}%
Clark, N., Macdonald, B., \& Kloo, I. (2020). A {B}ayesian adjusted
plus-minus analysis for the esport {Dota 2}. \emph{Journal of
Quantitative Analysis in Sports}, \emph{16}(4), 325--341.
\url{https://doi.org/10.1515/jqas-2019-0103}

\leavevmode\vadjust pre{\hypertarget{ref-cochran2017oxford}{}}%
Cochran, J., Bennett, J., \& Albert, J. (Eds.). (2017). \emph{The
{Oxford} anthology of statistics in sports, volume 1: 2000-2004} (p.
544). London: Oxford University Press.
\url{https://global.oup.com/academic/product/the-oxford-anthology-of-statistics-in-sports-9780198724926}

\leavevmode\vadjust pre{\hypertarget{ref-DeshpandeEvans2020}{}}%
Deshpande, S. K., \& Evans, K. (2020). Expected hypothetical completion
probability. \emph{Journal of Quantitative Analysis in Sports},
\emph{16}(2), 85--94. \url{https://doi.org/10.1515/jqas-2019-0050}

\leavevmode\vadjust pre{\hypertarget{ref-Deshpande2016}{}}%
Deshpande, S. K., \& Jensen, S. T. (2016). Estimating an {NBA} player's
impact on his team's chances of winning. \emph{Journal of Quantitative
Analysis in Sports}, \emph{12}(2).
\url{https://doi.org/10.1515/jqas-2015-0027}

\leavevmode\vadjust pre{\hypertarget{ref-deshpande2017hierarchical}{}}%
Deshpande, S. K., \& Wyner, A. (2017). A hierarchical bayesian model of
pitch framing. \emph{Journal of Quantitative Analysis in Sports},
\emph{13}(3), 95--112. \url{https://doi.org/10.1515/jqas-2017-0027}

\leavevmode\vadjust pre{\hypertarget{ref-dewan1993stats}{}}%
Dewan, J., \& Zminda, D. (1993). \emph{STATS basketball scoreboard
1993-1994} (p. 288). HarperPerennial.

\leavevmode\vadjust pre{\hypertarget{ref-R-retrosheet}{}}%
Douglas, C., \& Scriven, R. (2021). \emph{Retrosheet: Import
professional baseball data from retrosheet}.
\url{https://github.com/colindouglas/retrosheet}

\leavevmode\vadjust pre{\hypertarget{ref-R-sportyR}{}}%
Drucker, R. (2022). \emph{sportyR: Plot scaled ggplot representations of
sports playing surfaces}.
\url{https://github.com/sportsdataverse/sportyR}

\leavevmode\vadjust pre{\hypertarget{ref-Elo1978}{}}%
Elo, A. E. (1978). \emph{The rating of chessplayers, past and present}.
New York: Arco Publishing.

\leavevmode\vadjust pre{\hypertarget{ref-fernandez2021framework}{}}%
Fernandez, J., Bornn, L., \& Cervone, D. (2021). A framework for the
fine-grained evaluation of the instantaneous expected value of soccer
possessions. \emph{Machine Learning}, \emph{110}(6), 1389--1427.
\url{https://doi.org/10.1007/s10994-021-05989-6}

\leavevmode\vadjust pre{\hypertarget{ref-fide2022rating}{}}%
FIDE. (2022). \emph{Rating calculator}.
\url{https://ratings.fide.com/calc.phtml}

\leavevmode\vadjust pre{\hypertarget{ref-R-Lahman}{}}%
Friendly, M., Dalzell, C., Monkman, M., \& Murphy, D. (2022).
\emph{Lahman: Sean lahman baseball database}.
\url{https://CRAN.R-project.org/package=Lahman}

\leavevmode\vadjust pre{\hypertarget{ref-funt2022}{}}%
Funt, D. (2022). \emph{At {Sloan} sports conference, criticism mounts
over diversity, access}. The Washington Post.
\url{https://www.washingtonpost.com/sports/2022/06/13/sloan-sports-conference-diversity/}

\leavevmode\vadjust pre{\hypertarget{ref-gandar1988testing}{}}%
Gandar, J., Zuber, R., O'Brien, T., \& Russo, B. (1988). Testing
rationality in the point spread betting market. \emph{The Journal of
Finance}, \emph{43}(4), 995--1008.
\url{https://doi.org/10.1111/j.1540-6261.1988.tb02617.x}

\leavevmode\vadjust pre{\hypertarget{ref-gilani2022sportsdataverse}{}}%
Gilani, S. (2022). \emph{SportsDataverse}.
\url{https://sportsdataverse.org}

\leavevmode\vadjust pre{\hypertarget{ref-GlickmanSonas2015}{}}%
Glickman, M. E., \& Sonas, J. (2015). Introduction to the NCAA men's
basketball prediction methods issue. \emph{Journal of Quantitative
Analysis in Sports}, \emph{11}(1), 1--3.
\url{https://doi.org/10.1515/jqas-2015-0013}

\leavevmode\vadjust pre{\hypertarget{ref-Glickman1998}{}}%
Glickman, M. E., \& Stern, H. S. (1998). A state-space model for
national football league scores. \emph{Journal of the American
Statistical Association}, \emph{93}(441), 25--35.
\url{https://doi.org/10.1080/01621459.1998.10474084}

\leavevmode\vadjust pre{\hypertarget{ref-glickman2017estimating}{}}%
Glickman, M. E., \& Stern, H. S. (2017). Estimating team strength in the
NFL. In \emph{Handbook of statistical methods and analyses in sports}
(pp. 113--136). Boca Raton, FL: CRC Press.
\url{http://glicko.net/research/nfl-chapter.pdf}

\leavevmode\vadjust pre{\hypertarget{ref-goldner2012markov}{}}%
Goldner, K. (2012). A {M}arkov model of football: Using stochastic
processes to model a football drive. \emph{Journal of Quantitative
Analysis in Sports}, \emph{8}(1).
\url{https://doi.org/10.1515/1559-0410.1400}

\leavevmode\vadjust pre{\hypertarget{ref-goldner2017situational}{}}%
Goldner, K. (2017). Situational success: Evaluating decision-making in
football. In \emph{Handbook of statistical methods and analyses in
sports} (pp. 199--214). Boca Raton, FL: CRC Press.

\leavevmode\vadjust pre{\hypertarget{ref-Guan2022}{}}%
Guan, T., Nguyen, R., Cao, J., \& Swartz, T. (2022). In-game win
probabilities for the {National Rugby League}. \emph{The Annals of
Applied Statistics}, \emph{16}(1).
\url{https://doi.org/10.1214/21-aoas1514}

\leavevmode\vadjust pre{\hypertarget{ref-hamilton2011extension}{}}%
Hamilton, H. H. (2011). An extension of the pythagorean expectation for
association football. \emph{Journal of Quantitative Analysis in Sports},
\emph{7}(2). \url{https://doi.org/10.2202/1559-0410.1335}

\leavevmode\vadjust pre{\hypertarget{ref-healey2017new}{}}%
Healey, G. (2017). The new {M}oneyball: How ballpark sensors are
changing baseball. \emph{Proceedings of the IEEE}, \emph{105}(11),
1999--2002. \url{https://doi.org/10.1109/JPROC.2017.2756740}

\leavevmode\vadjust pre{\hypertarget{ref-healey2019bayesian}{}}%
Healey, G. (2019). A bayesian method for computing intrinsic pitch
values using kernel density and nonparametric regression estimates.
\emph{Journal of Quantitative Analysis in Sports}, \emph{15}(1), 59--74.
\url{https://doi.org/10.1515/jqas-2017-0058}

\leavevmode\vadjust pre{\hypertarget{ref-R-nflscrapR}{}}%
Horowitz, M., Yurko, R., Ventura, S., \& Dutta, R. (2020).
\emph{Nflscrap{R}: Compiling the NFL play-by-play API for easy use in
r}. \url{https://github.com/maksimhorowitz/nflscrapR}

\leavevmode\vadjust pre{\hypertarget{ref-hvattum2019comprehensive}{}}%
Hvattum, L. M. (2019). A comprehensive review of plus-minus ratings for
evaluating individual players in team sports. \emph{International
Journal of Computer Science in Sport}, \emph{18}(1).
\url{https://doi.org/10.2478/ijcss-2019-0001}

\leavevmode\vadjust pre{\hypertarget{ref-imbrogno2022computing}{}}%
Imbrogno, J., \& Bergman, D. (2022). Computing the number of winning NFL
survivor pool entries. \emph{The College Mathematics Journal},
\emph{53}(4), 282--291.
\url{https://doi.org/10.1080/07468342.2022.2099704}

\leavevmode\vadjust pre{\hypertarget{ref-james2003new}{}}%
James, B. (2003). \emph{The new {Bill James} historical baseball
abstract}. Free Press.

\leavevmode\vadjust pre{\hypertarget{ref-kaplan2001march}{}}%
Kaplan, E. H., \& Garstka, S. J. (2001). March madness and the office
pool. \emph{Management Science}, \emph{47}(3), 369--382.
\url{https://doi.org/10.1287/mnsc.47.3.369.9769}

\leavevmode\vadjust pre{\hypertarget{ref-koning2017rating}{}}%
Koning, R. H. (2017). Rating of team abilities in soccer. In
\emph{Handbook of statistical methods and analyses in sports} (pp.
371--388). Boca Raton, FL: CRC Press.

\leavevmode\vadjust pre{\hypertarget{ref-koopman2015dynamic}{}}%
Koopman, S. J., \& Lit, R. (2015). A dynamic bivariate poisson model for
analysing and forecasting match results in the english premier league.
\emph{Journal of the Royal Statistical Society: Series A (Statistics in
Society)}, \emph{178}(1), 167--186.
\url{https://doi.org/10.1111/rssa.12042}

\leavevmode\vadjust pre{\hypertarget{ref-kovalchik2016searching}{}}%
Kovalchik, S. A. (2016). Searching for the GOAT of tennis win
prediction. \emph{Journal of Quantitative Analysis in Sports},
\emph{12}(3), 127--138. \url{https://doi.org/10.1515/jqas-2015-0059}

\leavevmode\vadjust pre{\hypertarget{ref-kovalchik2019calibration}{}}%
Kovalchik, S. A., \& Reid, M. (2019). A calibration method with dynamic
updates for within-match forecasting of wins in tennis.
\emph{International Journal of Forecasting}, \emph{35}(2), 756--766.
\url{https://doi.org/10.1016/j.ijforecast.2017.11.008}

\leavevmode\vadjust pre{\hypertarget{ref-kumagai2021hockey}{}}%
Kumagai, B., Nahabedian, M., Châtel, T., \& Stokes, T. (2021).
\emph{Bayesian space-time models for expected possession added value}.
Hockey-Graphs.
\url{https://hockey-graphs.com/2021/07/06/bayesian-space-time-models-for-expected-possession-added-value-part-1-of-2/}

\leavevmode\vadjust pre{\hypertarget{ref-lacey1990estimation}{}}%
Lacey, N. J. (1990). An estimation of market efficiency in the {NFL}
point spread betting market. \emph{Applied Economics}, \emph{22}(1),
117--129. \url{https://doi.org/10.1080/00036849000000056}

\leavevmode\vadjust pre{\hypertarget{ref-R-chess}{}}%
Lente, C. (2020). \emph{Chess: Read, write, create and explore chess
games}. \url{https://github.com/curso-r/chess}

\leavevmode\vadjust pre{\hypertarget{ref-lewis2004moneyball}{}}%
Lewis, M. (2004). \emph{Moneyball: The art of winning an unfair game}
(p. 336). New York: WW Norton \& Company.

\leavevmode\vadjust pre{\hypertarget{ref-Lindsey1961}{}}%
Lindsey, G. R. (1961). The progress of the score during a baseball game.
\emph{Journal of the American Statistical Association}, \emph{56}(295),
703--728. \url{https://doi.org/10.1080/01621459.1961.10480656}

\leavevmode\vadjust pre{\hypertarget{ref-lindsey1963investigation}{}}%
Lindsey, G. R. (1963). An investigation of strategies in baseball.
\emph{Operations Research}, \emph{11}(4), 477--501.
\url{https://doi.org/10.1287/opre.11.4.477}

\leavevmode\vadjust pre{\hypertarget{ref-Lock2014}{}}%
Lock, D., \& Nettleton, D. (2014). Using random forests to estimate win
probability before each play of an {NFL} game. \emph{Journal of
Quantitative Analysis in Sports}, \emph{10}(2).
\url{https://doi.org/10.1515/jqas-2013-0100}

\leavevmode\vadjust pre{\hypertarget{ref-lopezDM}{}}%
Lopez, M. J. (2022). personal communication.

\leavevmode\vadjust pre{\hypertarget{ref-lopez2020bigger}{}}%
Lopez, M. J. (2020). Bigger data, better questions, and a return to
fourth down behavior: An introduction to a special issue on tracking
data in the national football league. \emph{Journal of Quantitative
Analysis in Sports}, \emph{16}(2), 73--79.
\url{https://doi.org/10.1515/jqas-2020-0057}

\leavevmode\vadjust pre{\hypertarget{ref-lopez2015building}{}}%
Lopez, M. J., \& Matthews, G. J. (2015). Building an NCAA men's
basketball predictive model and quantifying its success. \emph{Journal
of Quantitative Analysis in Sports}, \emph{11}(1), 5--12.
\url{https://doi.org/10.1515/jqas-2014-0058}

\leavevmode\vadjust pre{\hypertarget{ref-Lopez2018}{}}%
Lopez, M. J., Matthews, G. J., \& Baumer, B. S. (2018). How often does
the best team win? A unified approach to understanding randomness in
{North American} sport. \emph{The Annals of Applied Statistics},
\emph{12}(4). \url{https://doi.org/10.1214/18-aoas1165}

\leavevmode\vadjust pre{\hypertarget{ref-macdonald2012expected}{}}%
Macdonald, B. (2012). An expected goals model for evaluating NHL teams
and players. \emph{Proceedings of the 2012 MIT Sloan Sports Analytics
Conference}.
\url{https://assets.pubpub.org/mku181yp/ee0d61ed-af35-4b1f-ba86-71c216935690.pdf}

\leavevmode\vadjust pre{\hypertarget{ref-marchi2018analyzing}{}}%
Marchi, M., Albert, J., \& Baumer, B. S. (2018). \emph{Analyzing
baseball data with {R}} (2nd ed., p. 360). Boca Raton, FL: Chapman;
Hall/CRC Press. \url{https://doi.org/10.1201/9781351107099}

\leavevmode\vadjust pre{\hypertarget{ref-martin2018markov}{}}%
Martin, R., Timmons, L., \& Powell, M. (2018). A {M}arkov chain analysis
of {NFL} overtime rules. \emph{Journal of Sports Analytics},
\emph{4}(2), 95--105. \url{https://doi.org/10.3233/JSA-170198}

\leavevmode\vadjust pre{\hypertarget{ref-maymin2021smart}{}}%
Maymin, P. Z. (2021). Smart kills and worthless deaths: eSports
analytics for league of legends. \emph{Journal of Quantitative Analysis
in Sports}, \emph{17}(1), 11--27.
\url{https://doi.org/10.1515/jqas-2019-0096}

\leavevmode\vadjust pre{\hypertarget{ref-McFarlane2019}{}}%
McFarlane, P. (2019). Evaluating {NBA} end-of-game decision-making.
\emph{Journal of Sports Analytics}, \emph{5}(1), 17--22.
\url{https://doi.org/10.3233/jsa-180231}

\leavevmode\vadjust pre{\hypertarget{ref-metrick1996march}{}}%
Metrick, A. (1996). March madness? Strategic behavior in {NCAA}
basketball tournament betting pools. \emph{Journal of Economic Behavior
\& Organization}, \emph{30}(2), 159--172.
\url{https://doi.org/10.1016/S0167-2681(96)00855-4}

\leavevmode\vadjust pre{\hypertarget{ref-miller2007derivation}{}}%
Miller, S. J. (2007). A derivation of the pythagorean won-loss formula
in baseball. \emph{Chance}, \emph{20}(1), 40--48.
\url{https://doi.org/10.1080/09332480.2007.10722831}

\leavevmode\vadjust pre{\hypertarget{ref-Mills1970}{}}%
Mills, E. G., \& Mills, H. D. (1970). \emph{Player win averages: A
complete guide to winning baseball players}. The Harlan D. Mills
Collection. \url{https://trace.tennessee.edu/utk_harlan/6/}

\leavevmode\vadjust pre{\hypertarget{ref-nichols2014impact}{}}%
Nichols, M. W. (2014). The impact of visiting team travel on game
outcome and biases in {NFL} betting markets. \emph{Journal of Sports
Economics}, \emph{15}(1), 78--96.
\url{https://doi.org/10.1177/1527002512440580}

\leavevmode\vadjust pre{\hypertarget{ref-niemi2008contrarian}{}}%
Niemi, J. B., Carlin, B. P., \& Alexander, J. M. (2008). Contrarian
strategies for {NCAA} tournament pools: A cure for {M}arch madness?
\emph{Chance}, \emph{21}(1), 35--42.
\url{https://doi.org/10.1080/09332480.2008.10722884}

\leavevmode\vadjust pre{\hypertarget{ref-PaulWeinbach2014}{}}%
Paul, R. J., \& Weinbach, A. P. (2014). Market efficiency and behavioral
biases in the {WNBA} betting market. \emph{International Journal of
Financial Studies}, \emph{2}, 193--202.
\url{https://doi.org/10.3390/ijfs2020193}

\leavevmode\vadjust pre{\hypertarget{ref-pelechrinis2019}{}}%
Pelechrinis, K., Winston, W., Sagarin, J., \& Cabot, V. (2019).
Evaluating NFL plays: Expected points adjusted for schedule.
\emph{International Workshop on Machine Learning and Data Mining for
Sports Analytics}, \emph{11330}, 106--117.
\url{https://doi.org/10.1007/978-3-030-17274-9_9}

\leavevmode\vadjust pre{\hypertarget{ref-R-baseballr}{}}%
Petti, B., \& Gilani, S. (2022). \emph{Baseballr: Acquiring and
analyzing baseball data}.
\url{https://CRAN.R-project.org/package=baseballr}

\leavevmode\vadjust pre{\hypertarget{ref-R-base}{}}%
R Core Team. (2022). \emph{R: A language and environment for statistical
computing}. R Foundation for Statistical Computing.
\url{https://www.R-project.org/}

\leavevmode\vadjust pre{\hypertarget{ref-Reyers2021quarterback}{}}%
Reyers, M., \& Swartz, T. B. (2021). Quarterback evaluation in the
{N}ational {F}ootball {L}eague using tracking data. \emph{{AStA}
Advances in Statistical Analysis}.
\url{https://doi.org/10.1007/s10182-021-00406-8}

\leavevmode\vadjust pre{\hypertarget{ref-Romer2006}{}}%
Romer, D. (2006). Do firms maximize? Evidence from professional
football. \emph{Journal of Political Economy}, \emph{114}(2), 340--365.
\url{https://doi.org/10.1086/501171}

\leavevmode\vadjust pre{\hypertarget{ref-sauer1998economics}{}}%
Sauer, R. D. (1998). The economics of wagering markets. \emph{Journal of
Economic Literature}, \emph{36}(4), 2021--2064.
\url{https://www.jstor.org/stable/2565046}

\leavevmode\vadjust pre{\hypertarget{ref-Schuhmann2021nba3point}{}}%
Schuhmann, J. (2021). \emph{{NBA}'s 3-point revolution: {H}ow 1 shot is
changing the game}. NBA.com.
\url{https://www.nba.com/news/3-point-era-nba-75}

\leavevmode\vadjust pre{\hypertarget{ref-Schwarz2005numbers}{}}%
Schwarz, A. (2004). \emph{The numbers game: Baseball's lifelong
fascination with statistics}. New York: Thomas Dunne Books/St. Martin's
Press.

\leavevmode\vadjust pre{\hypertarget{ref-Sicilia2019DeepHoops}{}}%
Sicilia, A., Pelechrinis, K., \& Goldsberry, K. (2019, July).
{DeepHoops: Evaluating Micro-Actions in Basketball Using Deep Feature
Representations of Spatio-Temporal Data}. \emph{Proceedings of the 25th
{ACM} {SIGKDD} International Conference on Knowledge Discovery \& Data
Mining}. \url{https://doi.org/10.1145/3292500.3330719}

\leavevmode\vadjust pre{\hypertarget{ref-sidle2018using}{}}%
Sidle, G., \& Tran, H. (2018). Using multi-class classification methods
to predict baseball pitch types. \emph{Journal of Sports Analytics},
\emph{4}(1), 85--93. \url{https://doi.org/10.3233/JSA-170171}

\leavevmode\vadjust pre{\hypertarget{ref-R-pitchRx}{}}%
Sievert, C. (2015). \emph{pitchRx: Tools for harnessing MLBAM 'gameday'
data and visualizing pitchfx}. \url{http://cpsievert.github.com/pitchRx}

\leavevmode\vadjust pre{\hypertarget{ref-skinner2011scoring}{}}%
Skinner, B. (2011). Scoring strategies for the underdog: A general,
quantitative method for determining optimal sports strategies.
\emph{Journal of Quantitative Analysis in Sports}, \emph{7}(4).
\url{https://doi.org/10.2202/1559-0410.1364}

\leavevmode\vadjust pre{\hypertarget{ref-soebbing2013gamblers}{}}%
Soebbing, B. P., \& Humphreys, B. R. (2013). Do gamblers think that
teams tank? Evidence from the {NBA}. \emph{Contemporary Economic
Policy}, \emph{31}(2), 301--313.
\url{https://doi.org/10.1111/j.1465-7287.2011.00298.x}

\leavevmode\vadjust pre{\hypertarget{ref-spann2009sports}{}}%
Spann, M., \& Skiera, B. (2009). Sports forecasting: A comparison of the
forecast accuracy of prediction markets, betting odds and tipsters.
\emph{Journal of Forecasting}, \emph{28}(1), 55--72.
\url{https://doi.org/10.1002/for.1091}

\leavevmode\vadjust pre{\hypertarget{ref-Stern1994}{}}%
Stern, H. S. (1994). A brownian motion model for the progress of sports
scores. \emph{Journal of the American Statistical Association},
\emph{89}(427), 1128--1134.
\url{https://doi.org/10.1080/01621459.1994.10476851}

\leavevmode\vadjust pre{\hypertarget{ref-Tango2007book}{}}%
Tango, T. M., Lichtman, M. G., \& Dolphin, A. E. (2007). \emph{The book:
Playing the percentages in baseball}. Sterling, VA: Potomac Books.

\leavevmode\vadjust pre{\hypertarget{ref-R-BradleyTerry2}{}}%
Turner, H., \& Firth, D. (2020). \emph{BradleyTerry2: Bradley-terry
models}. \url{https://github.com/hturner/BradleyTerry2}

\leavevmode\vadjust pre{\hypertarget{ref-urschel2011}{}}%
Urschel, J., \& Zhuang, J. (2011). Are NFL coaches risk and loss averse?
Evidence from their use of kickoff strategies. \emph{Journal of
Quantitative Analysis in Sports}, \emph{7}(3), 14.
\url{https://doi.org/10.2202/1559-0410.1311}

\leavevmode\vadjust pre{\hypertarget{ref-White2002}{}}%
White, C., \& Berry, S. (2002). Tiered polychotomous regression: Ranking
NFL quarterbacks. \emph{The American Statistician}, \emph{56}(1),
10--21. \url{https://doi.org/10.1198/000313002753631312}

\leavevmode\vadjust pre{\hypertarget{ref-R-ggplot2}{}}%
Wickham, H., Chang, W., Henry, L., Pedersen, T. L., Takahashi, K.,
Wilke, C., Woo, K., Yutani, H., \& Dunnington, D. (2022). \emph{ggplot2:
Create elegant data visualisations using the grammar of graphics}.
\url{https://CRAN.R-project.org/package=ggplot2}

\leavevmode\vadjust pre{\hypertarget{ref-Winston2022}{}}%
Winston, W. L., Nestler, S., \& Pelechrinis, K. (2022).
\emph{Mathletics: How gamblers, managers, and fans use mathematics in
sports} (2nd ed.). Princeton, NJ: Princeton University Press.

\leavevmode\vadjust pre{\hypertarget{ref-Yam2019}{}}%
Yam, D. R., \& Lopez, M. J. (2019). What was lost? A causal estimate of
fourth down behavior in the national football league. \emph{Journal of
Sports Analytics}, \emph{5}(3), 153--167.
\url{https://doi.org/10.3233/jsa-190294}

\leavevmode\vadjust pre{\hypertarget{ref-Yurko2020going}{}}%
Yurko, R., Matano, F., Richardson, L. F., Granered, N., Pospisil, T.,
Pelechrinis, K., \& Ventura, S. L. (2020). Going deep: Models for
continuous-time within-play valuation of game outcomes in american
football with tracking data. \emph{Journal of Quantitative Analysis in
Sports}, \emph{16}(2), 163--182.
\url{https://doi.org/10.1515/jqas-2019-0056}

\leavevmode\vadjust pre{\hypertarget{ref-Yurko2019nflwar}{}}%
Yurko, R., Ventura, S., \& Horowitz, M. (2019). {nflWAR}: A reproducible
method for offensive player evaluation in football. \emph{Journal of
Quantitative Analysis in Sports}, \emph{15}(3), 163--183.
\url{https://doi.org/10.1515/jqas-2018-0010}

\leavevmode\vadjust pre{\hypertarget{ref-R-chessR}{}}%
Zivkovic, J. (2022). \emph{chessR: Functions to extract, clean and
analyse online chess game data}. \url{https://github.com/JaseZiv/chessR}

\end{CSLReferences}

\bibliographystyle{unsrt}
\bibliography{refs.bib, pkgs.bib}

\end{document}